\documentclass[11pt]{article}
\begin{document}
\title{The end-to-end distribution function
for a flexible chain with weak excluded-volume
interactions}

\author{A.D. Drozdov\footnote{
Phone: 972-86472146.
E-mail: aleksey@bgumail.bgu.ac.il}\\
Department of Chemical Engineering\\
Ben-Gurion University of the Negev\\
P.O. Box 653\\
Beer-Sheva 84105, Israel}
\date{}
\maketitle

\begin{abstract}
An explicit expression is derived for the distribution
function of end-to-end vectors and for the mean square
end-to-end distance of a flexible chain with
excluded-volume interactions.
The Hamiltonian for a flexible chain with weak
intra-chain interactions is determined by two small
parameters: the ratio $\varepsilon$
of the energy of interaction between segments
(within a sphere whose radius coincides with the
cut-off length for the potential) to the thermal energy,
and the ratio $\delta$ of the cut-off length to
the radius of gyration for a Gaussian chain.
Unlike conventional approaches grounded on the
mean-field evaluation of the end-to-end
distance, the Green function is found
explicitly (in the first approximation with respect
to $\varepsilon$).
It is demonstrated that (i)
the distribution function depends on $\varepsilon$
in a regular way, while its dependence on $\delta$
is singular, and (ii) the leading term in the
expression for the mean square end-to-end
distance linearly grows with $\varepsilon$ and
remains independent of $\delta$.
\end{abstract}
\vspace*{5 mm}

\noindent
{\bf Key-words:}
Flexible chains,
Excluded-volume interaction,
Path integral,
Distribution function
\vspace*{5 mm}

\noindent
{\bf PACS:} 05.90.+m, 82.35.Lr, 87.10.+e

\newpage

\section{Introduction}

This paper is concerned with the distribution function
of end-to-end vectors for a flexible chain with weak
excluded-volume interactions.
The effect of intra-chain interactions on the characteristic
size of ``real" flexible chains has attracted
substantial attention from the beginning of the 50s of
the past century, see early studies
\cite{Flo49}--\cite{KSR60}.
A noticeable progress has been reached in this area
during the past five decades, see \cite{Sch62}--\cite{Vil98},
to mention a few works,
and the results of investigation have been summarized
in a number of monographs
\cite{DE86}--\cite{Sch99}.
Nevertheless, the effect of segment interactions
on the statistics of macromolecules has remained
a subject of debate in the past five years.
This conclusion is confirmed by a large number
of publications, where this phenomenon
is discussed, see \cite{SS99}--\cite{SE04},
and it may be ascribed to the importance of
steric interactions for the description of
such phenomena as
(i) protein folding and denaturation \cite{CG04},
(ii) unzipping of DNA molecules \cite{LLH04},
(iii) force--stretch relations for long chains
at relatively small elongations \cite{Lam02},
and (iv) transport of flexible chains through
narrow pores \cite{KM04}.
The account for intra-chain interactions
(i) provides a consistent description of
the dependence of intrinsic viscosity on
molecular weight in dilute polymer solutions
\cite{FTN77,NF78,KTE96},
(ii) leads to an adequate characterization
of time-dependent shear stresses in polymer fluids
\cite{GW92,Car95,KTE96,PR00,AK02,BW03,CT04},
(iii) allows transport coefficients in polymer
melts to be correctly estimated \cite{AK02},
and (iv) predicts reinforcements of interfaces between
incompatible solvents by random copolymers \cite{DTC00}.
Interactions between segments are assumed to be
responsible for
(i) swelling of polymer gels \cite{MJ69,Vil98},
(ii) mechanical \cite{BW03,CT04} and optical \cite{KTE96}
anisotropy of polymer solutions,
(iii) structure of polyelectrolyte chains
\cite{OH78,RM92,GCM01},
(iv) stability of globular proteins \cite{ZH96,CG04},
etc.

There are two conventional ways to describe
excluded-volume interactions in a flexible chain.
According to the first, a freely jointed chain
is thought of as a set of rigid segments with
given length linked in sequel, whereas the intra-chain
interactions do not permit each pair of segments
to occupy the same place.
This model implies that the distribution function for
end-to-end vectors coincides with that for a
self-avoiding random walk.
Although an explicit expression for the latter function
is unknown, it is believed that the main term in the
expansion of the distribution function
for a long chain (with the number of segments $N\gg 1$)
is described by either a stretched exponential function
or a stretched exponential function multiplied by some power
of the end-to-end distance
\cite{MM71,Clo74}.
Results of numerical simulation confirm these hypotheses,
but appropriate exponents are not precisely known
\cite{BC91}.

According to the other method, a chain is treated
as a sequence of beads bridged by (entropic) elastic
springs \cite{DE86}.
The Hamiltonian of the system equals the sum of
the conventional Hamiltonian for a Gaussian chain
and the excluded-volume functional, where the energy
of repulsive interactions between
segments is approximated by the Dirac delta-function.
To assess the effect of steric interactions on
the mean square end-to-end distance, it is presumed
that the intensity of interactions is small compared
to thermal energy.

By applying various approximate techniques, the
average end-to-end distance was evaluated in a number
of studies (to the best of our knowledge, only a few
attempts were undertaken to determine the
distribution function of end-to-end vectors \cite{Sch62,BCR91}).
It is widely accepted \cite{DE86,CJ90} that (i) the mean square
end-to-end distance $B$ of a flexible chain
with segment interactions is determined by
some small dimensionless parameter $\epsilon$ (which
quantity coincides with the product $\varepsilon\delta$
introduced later),
(ii) the ratio $(B/b)^{2}$, where $b$ is the mean
square end-to-end distance for a Gaussian chain,
may be expanded into a Taylor series in $\epsilon$,
and (iii) the leading term (proportional to $\epsilon$)
in this expansion is positive, which means that
excluded-volume interactions are always repulsive.
The common feature of the mathematical methods
applied to develop these results
was that all of them were grounded on the mean-field
approach (the propagator for an appropriate path
integral was evaluated on the classical path only).

The objective of the present study is to develop
an explicit expression for the distribution function
of end-to-end vectors by evaluating the functional
integral on (more or less) arbitrary paths
(conformations of a chain).
Our main result is that conclusions (i) to (iii) appear
to be questionable.
It is demonstrated that the Green function for a
flexible chain with excluded-volume interactions
is characterized by two small parameters,
one of which, $\varepsilon$, describes the intensity
of intra-chain interactions per small volume
associated with the segment length,
while the other, $\delta$, determines the ratio of
the segment length to the average end-to-end distance
for a Gaussian chain.
The difference between these parameters is that
the dependence of the distribution function on
$\varepsilon$ is regular, whereas its dependence
on $\delta$ is singular.
It is found that the leading terms in the expression
for the mean square end-to-end distance is proportional
to $\varepsilon$, the scaling that substantially
differs from the results of previous works.
The growth of the average size of a chain with the
intensity of segment interactions is ensured when
the excluded-volume potential is included into
the Hamiltonian with the sign different from the
conventional one.

The exposition is organized as follows.
A Hamiltonian for a flexible chain with
excluded-volume interactions is discussed in
Section 2.
This Hamiltonian is calculated on the classical path
in Section 3.
Mean-field approximations for the Green
function and the end-to-end distance are given
in Section 4.
A perturbed Hamiltonian is determined in Section 5.
Our analysis is based on the conventional
approach, according to which the excluded-volume
potential is expanded in a Taylor series
in the vicinity of the classical path
up to quadratic terms (a similar
technique was recently employed in \cite{ZRG03}
for the screened Coulomb potential).
A leading term (with respect to small parameters)
in the expression for the Green function is
determined in Section 6.
The mean square end-to-end distance $B$ is
evaluated in Section 7 for weak ($\varepsilon\delta^{-2}\ll 1$)
and in Section 8 for moderately strong (arbitrary
values of this ratio) interactions.
Some concluding remarks are formulated in Section 9.
Mathematical derivations are presented in Appendices A to F.

\section{Formulation of the problem}

A flexible chain is treated as a curve with
``length" $L$ in a three-dimensional space.
An arbitrary configuration of the chain
is determined by the function ${\bf r}(s)$,
where ${\bf r}$ denotes radius vector and $s\in [0,L]$.
For definiteness, we assume the end $s=0$
to be fixed at the origin,
${\bf r}(0)={\bf 0}$, and the end $s=L$ to be
free.
An ``internal structure" of the curve is characterized
by a segment length $b_{0}$ and the number of
segments $N\gg 1$, which are connected with
length $L$ by the formula
$L=b_{0} N$.
In the absence of segment interactions,
the Hamiltonian
\begin{equation}
H_{0}({\bf r})=\frac{3k_{\rm B}T}{2b_{0}}\int_{0}^{L}
\Bigl (\frac{d{\bf r}}{ds}(s)\Bigr )^{2} ds
\end{equation}
is ascribed to the chain.
Here $k_{\rm B}$ is Boltmann's constant,
and $T$ is the absolute temperature.
The energy of intra-chain interactions is described
by the functional
\begin{equation}
\bar{V}({\bf r})=\frac{v_{0}}{2L^{2}}
\int_{0}^{L} ds\int_{0}^{L}
V\Bigl ({\bf r}(s)-{\bf r}(s^{\prime})
\Bigr ) ds^{\prime},
\end{equation}
where $V({\bf r})$ is a dimensionless potential
energy of interactions,
and $v_{0}$ is their intensity (which is assumed
to be positive for repulsive and negative for
attractive interactions).
We confine ourselves to isotropic functions
$V({\bf r})=V_{\ast}(r)$ with $r=|{\bf r}|$.
The entire Hamiltonian of the chain reads
\begin{equation}
H=H_{0}+\bar{V},
\end{equation}
which implies that
\begin{equation}
H({\bf r})=k_{\rm B}T\biggl [
\frac{3}{2b_{0}}\int_{0}^{L}
\Bigl (\frac{d{\bf r}}{ds}(s)\Bigr )^{2} ds
+\frac{v_{0}}{2k_{\rm B}T L^{2}} \int_{0}^{L} ds
\int_{0}^{L} V\Bigl ({\bf r}(s)-{\bf r}(s^{\prime})
\Bigr ) ds^{\prime} \biggr ].
\end{equation}
Although Eq. (3) is employed in a number of studies,
see \cite{DE86} and the references therein,
the sign of the second term in this formula
seems questionable,
because $\bar{V}$ may be considered as a work of some
fictitious external field that prevents segments from
being located at the same positions.
The latter implies that the contribution of $\bar{V}$
into $H$ should be negative (as, for example, that
of the work of an external force at extension of a
chain).
We postpone a discussion of the sign of $\bar{V}$
to Section 7, and suppose at this stage that $v_{0}$
may have an arbitrary sign.

The Green function (propagator) for a chain $G({\bf Q})$
is given by
\begin{equation}
G({\bf Q})=\int_{{\bf r}(0)={\bf 0}}^{{\bf r}(L)={\bf Q}}
\exp \Bigl [-\frac{H({\bf r})}{k_{\rm B}T}\Bigr ]
{\cal D}({\bf r}(s)),
\end{equation}
where the path integral is calculated over all curves
${\bf r}(s)$ that obey the boundary conditions
\begin{equation}
{\bf r}(0)={\bf 0},
\qquad
{\bf r}(L)={\bf Q}.
\end{equation}
The aim of this study is to develop an explicit
expression for the Green function of a chain
with relatively weak segment interactions,
when the last term in Eq. (4) is ``small" compared
to the first.
The derivation is performed for an arbitrary
function $V({\bf r})$, but our specific
interest is in the excluded-volume potential
$V({\bf r})=\hat{\delta}_{\rm D}({\bf r})$,
where $\delta_{\rm D}({\bf r})$ denotes
the Dirac delta-function,
and $\hat{\delta}_{\rm D}({\bf r})$ stands for its
regularization after cut-off at small distances.
To provide a rigorous definition,
we introduce the Fourier transform of $V({\bf r})$,
\begin{equation}
U({\bf k})=\int V({\bf r})
\exp (\imath {\bf k}\cdot {\bf r})d{\bf r}
\end{equation}
with the inverse
\begin{equation}
V({\bf r})=\frac{1}{(2\pi)^{3}}
\int U({\bf k}) \exp (
-\imath {\bf k}\cdot {\bf r})d{\bf k},
\end{equation}
where the dot stands for inner product.
Equation (7) implies that for an isotropic function
$V({\bf r})$, the function $U({\bf k})$ is isotropic
as well,
\[
U({\bf k})=U_{\ast}(k),
\qquad
U_{\ast}(k)=\frac{4\pi}{k}\int_{0}^{\infty}
V_{\ast}(r)\sin (kr)r dr.
\]
The Fourier transform of the Dirac delta-function
$\delta_{\rm D}({\bf r})$ reads
$U_{{\rm D}\ast}(k)=1$.
To avoid divergence of the integral in Eq. (4),
this function is cut off at an appropriate
length-scale $b_{\ast}=k_{\ast}^{-1}$,
\begin{equation}
\hat{U}_{{\rm D} \ast}(k)=\left \{ \begin{array}{ll}
1, & k\leq k_{\ast},
\\
0, & k> k_{\ast},
\end{array}\right .
\end{equation}
and the potential $\hat{\delta}_{\rm D}({\bf r})$
is defined as the inverse Fourier transform of
$\hat{U}_{{\rm D}\ast}(k)$.
To ensure that the potential $\hat{U}_{{\rm D}\ast}$
describes intra-chain interactions properly, we
assume the length-scale of cut-off $b_{\ast}$ to be
substantially lower than the mean-square end-to-end
distance for a Gaussian chain $b=\sqrt{b_{0}L}$,
\begin{equation}
\delta=\frac{b_{\ast}}{b}\ll 1.
\end{equation}

\section{Weak excluded-volume interactions}

To characterize the ``smallness" of the potential
$\bar{V}$, we, first, introduce a test function
${\bf r}_{0}$ (the classical path) that minimizes
functional (1) under conditions (6).
Then the values of $H_{0}$ and $\bar{V}$
are determined on the curve ${\bf r}_{0}(s)$,
and the dimensionless parameter $\varepsilon$ is
chosen from the condition that $\bar{V}({\bf r}_{0})$
is small compared with $H_{0}({\bf r}_{0})$
for any properly scaled end-to-end vector ${\bf Q}$.
We introduce a Cartesian coordinate frame $\{ x,y,z \}$,
whose $z$ axis is directed along the vector ${\bf Q}$,
and present the minimization problem for the non-perturbed
Hamiltonian $H_{0}$ as follows:
\begin{eqnarray*}
&& \min \int_{0}^{L} \Bigl [ \Bigl (\frac{dx}{ds}(s)\Bigr )^{2}
+\Bigl (\frac{dy}{ds}(s)\Bigr )^{2}
+\Bigl (\frac{dz}{ds}(s)\Bigr )^{2}\Bigr ] ds,
\nonumber\\
&& x(0)=y(0)=z(0)=0,
\quad
x(L)=Q,
\quad
y(L)=z(L)=0.
\end{eqnarray*}
The unique solution of this problem is given by
\begin{equation}
{\bf r}_{0}(s)={\bf Q}\frac{s}{L}.
\end{equation}
It follows from Eqs. (1) and (11) that
\begin{equation}
H_{0}({\bf r}_{0})
=\frac{3}{2}k_{\rm B}T \tilde{Q}^{2},
\qquad
\tilde{Q}=\frac{Q}{b}.
\end{equation}
Equation (8) implies that
\[
V\Bigl ({\bf r}(s)-{\bf r}(s^{\prime})\Bigr )
=\frac{1}{(2\pi)^{3}}\int U({\bf k})
\exp \Bigl [-\imath {\bf k}\cdot \Bigl (
{\bf r}(s)-{\bf r}(s^{\prime})\Bigr )\Bigr ]
d{\bf k} .
\]
Substitution of this expression into Eq. (2)
results in
\begin{equation}
\bar{V}({\bf r})=\frac{v_{0}}{2L^{2}(2\pi)^{3}}
\int U({\bf k})d{\bf k}
\int_{0}^{L} ds\int_{0}^{L}
\exp \Bigl [-\imath {\bf k}\cdot \Bigl (
{\bf r}(s)-{\bf r}(s^{\prime})\Bigr )\Bigr ]
ds^{\prime}.
\end{equation}
Combining Eqs. (11) and (13), we find that
\begin{equation}
\bar{V}({\bf r}_{0})=\frac{v_{0}}{2L^{2}(2\pi)^{3}}
\int U({\bf k})d{\bf k}
\int_{0}^{L} ds\int_{0}^{L}
\exp \Bigl (-\imath {\bf k}\cdot {\bf Q}
\frac{s-s^{\prime}}{L} \Bigr ) ds^{\prime}.
\end{equation}
Simple algebra implies that (Appendix A)
\begin{equation}
\bar{V}({\bf r}_{0})
= \frac{v_{0}}{2(\pi b\tilde{Q})^{2}} \int_{0}^{1}
\frac{dx}{x^{2}} \int_{0}^{\infty}
U_{\ast}(k) \Bigl (1-\cos (kb\tilde{Q} x)\Bigr )dk.
\end{equation}
Formula (15) determines the energy of intra-chain
interactions for an arbitrary function $U_{\ast}(k)$.
In particular, for the excluded-volume potential (9),
this equality reads
\begin{equation}
\bar{V}({\bf r}_{0})
= \frac{v_{0}}{2(\pi b\tilde{Q})^{2}} \int_{0}^{1}
\frac{dx}{x^{2}} \int_{0}^{k_{\ast}}
\Bigl (1-\cos (kb\tilde{Q} x)\Bigr )dk.
\end{equation}
Calculation of the integral (Appendix A) results in
\begin{equation}
\bar{V}({\bf r}_{0})
= \frac{v_{0}k_{\ast}^{3}}{4\pi^{2}}A(k_{\ast}b\tilde{Q}),
\end{equation}
where
\begin{equation}
A(x)= \frac{1}{x^{3}}
\int_{0}^{x} (x^{2}-z^{2})\frac{1-\cos z}{z^{2}}d z.
\end{equation}
The function $A(x)$ is plotted in Figure 1,
which shows that $A$ monotonically decreases with $x$.
The limits of $A(x)$ as $x\to 0$ and
$x\to\infty$ are given by (Appendix A)
\begin{equation}
\lim_{x\to 0}A(x)=\frac{1}{3},
\quad
\lim_{x\to \infty} A(x)=0.
\end{equation}
It follows from Eqs. (12) and (17) that the smallness
of the functional $\bar{V}$
compared with the non-perturbed Hamiltonian $H_{0}$
is tantamount to that of the pre-factor
\[
\frac{v_{0}k_{\ast}^{3}}{4\pi^{2}}
=\frac{v_{0}}{3\pi\nu}
\]
compared with thermal energy $k_{\rm B}T$.
Here $\nu=\frac{4}{3} \pi b_{\ast}^{3}$ is the
volume in which interactions between segments
are taken into account.
Introducing the parameter $\varepsilon$ by
\begin{equation}
\varepsilon=\frac{v_{0}}{k_{\rm B}T\nu},
\end{equation}
we conclude that the condition $|\varepsilon|\ll 1$ means
that the energy of excluded-volume interactions at the
length-scale of cut-off is small compared with
thermal energy.
Substitution of expressions (12), (17) and (20) into
Eq. (4) results in
\begin{equation}
H({\bf r}_{0})=k_{\rm B}T\Bigl [ \frac{3}{2}\tilde{Q}^{2}
+\frac{\varepsilon}{3\pi} A(\frac{\tilde{Q}}{\delta})
\Bigr ].
\end{equation}
According to Eq. (21), in the mean-field approximation,
the functional $H$ is determined by two dimensionless
quantities, $\varepsilon$ and $\delta$.
The former characterizes the smallness of excluded-volume
interactions, whereas the latter describes the length-scale
where these interactions are important.
The parameter $\varepsilon$ is located in the numerator,
which means that it is responsible for regular
perturbations of $H_{0}$,
while $\delta$ stands in the denominator,
which implies that it describes singular perturbations.
This distinguishes the present approach from previous
studies, where one small parameter was introduced,
and its effect was assumed to be regular in the sense
that the Green function was expanded into a Taylor
series with respect to this quantity.

Before proceeding with the calculation of
the Hamiltonian $H$ on an arbitrary curve ${\bf r}(s)$,
it is instructive to evaluate the Green function
$G({\bf Q})$ and its second moment on the classical path
${\bf r}_{0}(s)$.

\section{Mean-field approximation of the Green function}

Our aim is to find an approximation of the mean
square end-to-end distance $B$ when the set of
admissible curves $\{ {\bf r}(s) \}$ contains
the classical path ${\bf r}_{0}(s)$ only.
Substitution of expression (21) into Eq. (5)
results in
\begin{equation}
G({\bf Q})=C\exp \Bigl [-\Bigl (\frac{3}{2}\tilde{Q}^{2}
+\frac{\varepsilon}{3\pi} A(\frac{\tilde{Q}}{\delta})
\Bigr )\Bigr ],
\end{equation}
where the pre-factor $C$ characterizes the measure
${\cal D}({\bf r}(s))$ of the function ${\bf r}_{0}(s)$.
To determine $C$, we treat the Green function
as the distribution function of end-to-end vectors ${\bf Q}$
that obeys the normalization condition
\[
\int G({\bf Q}) d{\bf Q}=1.
\]
In the spherical coordinate frame $\{ Q,\phi,\theta \}$
whose $z$ vector is directed along the vector ${\bf Q}$,
this equality reads
\begin{equation}
4\pi \int_{0}^{\infty} G(Q) Q^{2} dQ=1.
\end{equation}
Substituting expression (22) into Eq. (23) and
setting $Q=b\tilde{Q}$, we find that
\[
4\pi b^{3} C\int_{0}^{\infty}
\exp \Bigl [-\Bigl (\frac{3\tilde{Q}^{2}}{2}
+\frac{\varepsilon}{3\pi} A(\frac{\tilde{Q}}{\delta})
\Bigr )\Bigr ]\tilde{Q}^{2} d\tilde{Q}=1.
\]
Expanding the function under the integral into the
Taylor series (this is possible because the function
$A(x)$ is uniformly bounded) and disregarding terms
beyond the first order of smallness with respect
to $\varepsilon$, we obtain
\[
4\pi b^{3} C\int_{0}^{\infty}
\exp \Bigl (-\frac{3\tilde{Q}^{2}}{2}\Bigr )
\Bigl [ 1- \frac{\varepsilon}{3\pi}
A(\frac{\tilde{Q}}{\delta})\Bigr ]
\tilde{Q}^{2} d\tilde{Q}=1.
\]
It follows from this equality that with the
required level of accuracy,
\begin{equation}
C=\Bigl (\frac{3}{2\pi b^{2}}\Bigr )^{\frac{3}{2}}
\Bigl (1+\varepsilon \sqrt{\frac{6}{\pi^{3}}}
C_{1}\Bigr ),
\end{equation}
where
\begin{equation}
C_{1}=\int_{0}^{\infty} A(\frac{\tilde{Q}}{\delta})
\exp \Bigl (-\frac{3\tilde{Q}^{2}}{2}\Bigr )
\tilde{Q}^{2} d\tilde{Q}.
\end{equation}
The leading term in the expression for $C_{1}$ reads
(Appendix B)
\begin{equation}
C_{1}=\frac{\pi \delta}{6}.
\end{equation}
The mean square end-to-end distance of a chain
is given by
\[
B^{2}=\int Q^{2} G({\bf Q}) d{\bf Q}.
\]
In the spherical coordinates $\{ Q, \phi,\theta \}$,
this equality is presented in the form
\begin{equation}
B^{2}=4\pi \int_{0}^{\infty} G(Q) Q^{4} d Q.
\end{equation}
Substituting expression (22) into Eq. (27),
introducing the new variable $\tilde{Q}=Q/b$,
and neglecting terms beyond the first order
of smallness with respect to $\varepsilon$,
we find that
\[
B^{2}=4\pi b^{5}C \int_{0}^{\infty}
\exp \Bigl (-\frac{3\tilde{Q}^{2}}{2}\Bigr )
\Bigl [ 1-\frac{\varepsilon}{3\pi}
A(\frac{\tilde{Q}}{\delta})
\Bigr ]\tilde{Q}^{4} d\tilde{Q}.
\]
This equality implies that with the required level
of accuracy,
\begin{equation}
\Bigl (\frac{B}{b}\Bigr)^{2}
=C \Bigl (\frac{2\pi b^{2}}{3}\Bigr )^{\frac{3}{2}}
\Bigl (1-\varepsilon \sqrt{\frac{6}{\pi^{3}}}
B_{1}\Bigr ),
\end{equation}
where
\begin{equation}
B_{1} = \int_{0}^{\infty}
A(\frac{\tilde{Q}}{\delta})
\exp \Bigl (-\frac{3\tilde{Q}^{2}}{2} \Bigr )
\tilde{Q}^{4} d\tilde{Q} .
\end{equation}
The leading term in the expression for $B_{1}$
reads (Appendix B)
\begin{equation}
B_{1}=\frac{\pi\delta}{9}.
\end{equation}
It follows from Eqs. (24), (26), (28) and (30) that
\begin{equation}
\Bigl (\frac{B}{b}\Bigr)^{2}
=1+\varepsilon \sqrt{\frac{6}{\pi^{3}}}(C_{1}-B_{1})
=1+\frac{\varepsilon\delta}{9}\sqrt{\frac{3}{2\pi}}.
\end{equation}
Equation (31) implies that at $v_{0}>0$,
the mean square end-to-end distance of a flexible
chain with excluded-volume
interactions exceeds that of a Gaussian chain.
At first sight, this conclusion confirms that
the sign of the contribution of $\bar{V}$ into
the Hamiltonian $H$ is chosen correctly in Eq. (3).
Setting $b_{\ast}=b_{0}$ (the cut-off of the potential
$U_{\ast}$ occurs at the segment length)
and using Eqs. (10) and (20), we find that
\[
\Bigl (\frac{B}{b}\Bigr)^{2}
=1+\frac{v_{0}}{\sqrt{6\pi}k_{\rm B}T \nu_{\rm tube}},
\]
where $\nu_{\rm tube}=4 \pi b_{0}^{2}b$ denotes
volume of the characteristic tube around a chain
(a circular cylinder whose radius coincides
with the segment length and whose length equals
the mean square end-to-end distance).
Our aim now is to demonstrate that Eq. (31) does
not capture the main contribution into the
asymptotic expression for the average
end-to-end distance.
To prove this fact, we derive an explicit expression
for the Hamiltonian $H$ that accounts for second
order terms with respect to admissible fluctuations
from the classical path ${\bf r}_{0}(s)$.

\section{Perturbations of the Hamiltonian}

In accord with Eq. (6), the function ${\bf r}(s)$
is given by
\begin{equation}
{\bf r}(s)={\bf Q}\frac{s}{L}+{\bf R}(s),
\end{equation}
where the function ${\bf R}(s)$ satisfies the
boundary conditions
\begin{equation}
{\bf R}(0)={\bf 0},
\qquad
{\bf R}(L)={\bf 0}.
\end{equation}
To simplify the analysis, we neglect longitudinal
fluctuations and expand the function ${\bf R}(s)$
that describes transverse fluctuations into the
Fourier series
\begin{equation}
{\bf R}(s)=\sum_{m=1}^{\infty} X_{m}\sin\frac{\pi m s}{L}
{\bf e}_{1}
+\sum_{m=1}^{\infty} Y_{m}\sin\frac{\pi m s}{L}
{\bf e}_{2},
\end{equation}
where $X_{m}$, $Y_{m}$ are arbitrary coefficients,
and ${\bf e}_{k}$ ($k=1,2,3$) are unit vectors of
a Cartesian coordinate frame, whose ${\bf e}_{3}$
vector is directed along the vector ${\bf Q}$.
Any function ${\bf R}(s)$ obeying Eq. (34) satisfies
also boundary conditions (33).
We substitute expressions (32) and (34) into Eqs. (1)
and (2) and, after some algebra, find that (Appendix C)
\begin{eqnarray}
H_{0}({\bf r}) &=& H_{0}({\bf r}_{0})
+\frac{3\pi^{2}k_{\rm B}T}{4b^{2}}
\sum_{m=1}^{\infty} m^{2}(X_{m}^{2} +Y_{m}^{2}) ,
\nonumber\\
\bar{V}({\bf r}) &=& \bar{V}({\bf r}_{0})
-\frac{v_{0}}{(4\pi)^{2}}\sum_{m,n=1}^{\infty}
P_{mn}(X_{m}X_{n}+Y_{m}Y_{n}),
\end{eqnarray}
where
\begin{eqnarray}
P_{mn} &=& \int_{0}^{\infty} U_{\ast}(k) k^{4} dk
\int_{-1}^{1}(1-x^{2})dx
\biggl \{ F(k Qx )
\nonumber\\
&&\times
\Bigl [ \cos\frac{\pi(m-n)}{2}
\Bigl ( F(kQx+\pi(m-n))+F(kQx-\pi (m-n))\Bigr )
\nonumber\\
&&-\cos\frac{\pi(m+n)}{2}
\Bigl ( F(kQx+\pi(m+n))+F(kQx-\pi (m+n))\Bigr )\Bigr ]
\nonumber\\
&&- \cos\frac{\pi m}{2}\cos\frac{\pi n}{2}
\Bigl [ F(kQx+\pi m)- F(kQx-\pi m)\Bigr ]
\nonumber\\
&&\times
\Bigl [ F(kQx+\pi n)- F(kQx-\pi n)\Bigr ]
-\sin\frac{\pi m}{2}\sin\frac{\pi n}{2}
\Bigl [ F(kQx+\pi m)
\nonumber\\
&&+ F(kQx-\pi m) \Bigr ]
\Bigl [ F(kQx+\pi n)+F(kQx-\pi n) \Bigr ]\biggr \},
\end{eqnarray}
and the function $F(z)$ reads
\begin{equation}
F(z)=\frac{1}{z} \sin \frac{z}{2}.
\end{equation}
It follows from Eqs. (4) and (35) that with the
required level of accuracy, the perturbed
Hamiltonian $H$ is given by
\begin{equation}
H({\bf r})=H({\bf r}_{0})+k_{\rm B}T \biggl [
\frac{3\pi^{2}}{4b^{2}}\sum_{m=1}^{\infty}
m^{2}(X_{m}^{2}+Y_{m}^{2})
-\frac{\varepsilon \nu}{(4\pi)^{2}}\sum_{m,n=1}^{\infty}
P_{mn}(X_{m}X_{n}+Y_{m}Y_{n})\biggr ].
\end{equation}
Our purpose now is to substitute this expression into
Eq. (5) and to calculate the path integral.

\section{The Green function}

Combining Eqs. (5) and (38), we find that
\begin{eqnarray}
G({\bf Q}) &=& \exp \Bigl (
-\frac{H({\bf r}_{0})}{k_{\rm B}T}\Bigr )
\int_{{\bf R}(0)={\bf 0}}^{{\bf R}(L)={\bf 0}}
\exp \biggl \{
-\frac{1}{2} \biggl [ \frac{3\pi^{2}}{2b^{2}}
\sum_{m=1}^{\infty} m^{2}(X_{m}^{2}+Y_{m}^{2})
\nonumber\\
&& -\frac{\varepsilon \nu}{8\pi^{2}}
\sum_{m,n=1}^{\infty}
P_{mn}(X_{m}X_{n}+Y_{m}Y_{n})\biggr ] \biggr \}
{\cal D}({\bf R}(s)).
\end{eqnarray}
Bearing in mind Eq. (21) and using the matrix
presentation of the functional integral, we obtain
\[
G({\bf Q})=C \exp \Bigl [-\Bigl (\frac{3\tilde{Q}^{2}}{2}
+\frac{\varepsilon}{3\pi} A(\frac{\tilde{Q}}{\delta})
\Bigr )\Bigr ]
\Lambda^{2}({\bf Q}),
\qquad
\Lambda({\bf Q})=\int \exp \Bigl [-\frac{1}{2} {\bf X}
\cdot ({\bf A}+\varepsilon {\bf B})\cdot {\bf X} \Bigr ]
d{\bf X},
\]
where $C$ is a constant associated with transition
from the measure ${\cal D}({\bf R}(s))$ to the measure
$d{\bf X}$ (this quantity will be found from the
normalization condition for the Green function).
Here ${\bf X}$ is the vector with components $X_{m}$,
${\bf A}$ is the diagonal matrix with components
$A_{mm}=3\pi^{2}m^{2}/(2b^{2})$,
and ${\bf B}$ is the matrix with components
$B_{mn}=-\nu P_{mn}/(8\pi^{2})$.
Calculation of the Gaussian integral implies that
\begin{equation}
G({\bf Q})=C \exp \Bigl [-\Bigl (\frac{3\tilde{Q}^{2}}{2}
+\frac{\varepsilon}{3\pi} A(\frac{\tilde{Q}}{\delta})
\Bigr )\Bigr ]
\frac{\det {\bf A}}{\det({\bf A}+\varepsilon {\bf B})}.
\end{equation}
In the first approximation with respect to
$\varepsilon$,
\begin{equation}
\det ({\bf A}+\varepsilon {\bf B})=\det {\bf A}
\Bigl (1+\varepsilon \sum_{m=1}^{\infty}
\frac{B_{mm}}{A_{mm}}\Bigr ).
\end{equation}
It follows from Eqs. (40) and (41) that
\begin{equation}
G({\bf Q})=C \exp \Bigl [-\Bigl (\frac{3\tilde{Q}^{2}}{2}
+\frac{\varepsilon}{3\pi} A(\tilde{Q})\Bigr )\Bigr ]
\Bigl (1+\frac{\varepsilon\nu b^{2}}{12\pi^{4}}S \Bigr )^{-1},
\end{equation}
where
\begin{equation}
S=-\sum_{m=1}^{\infty} \frac{P_{mm}}{m^{2}} .
\end{equation}
Substitution of expression (36) into Eq. (43)
implies that (Appendix D)
\begin{equation}
S = \int_{0}^{\infty} U_{\ast}(k) k^{4} dk
\int_{-1}^{1} (1-x^{2}) S_{0}(kQx) dx,
\end{equation}
where
\begin{eqnarray}
S_{0}(z) &=& F(z)s_{1}(z)+s_{2}(z)-2s_{3}(z)
-\frac{\pi^{2}}{3}F^{2}(z),
\nonumber\\
s_{1}(z) &=& \sum_{m=1}^{\infty} \frac{(-1)^{m}}{m^{2}}
\Bigl [ F(z+2\pi m)+F(z-2\pi m) \Bigr ],
\nonumber\\
s_{2}(z) &=& \sum_{m=1}^{\infty} \frac{1}{m^{2}}\Bigl [
F^{2}(z+\pi m)+F^{2}(z-\pi m)\Bigr ],
\nonumber\\
s_{3}(z) &=& \sum_{m=1}^{\infty}  \frac{(-1)^{m}}{m^{2}}
F(z+\pi m)F(z-\pi m).
\end{eqnarray}
The infinite sums are calculated in Appendix D,
where it is shown that
\begin{equation}
S_{0}(z)= \frac{\pi^{2}}{2z^{2}}\alpha(z),
\qquad
\alpha(z)=1+2\frac{\sin z}{z}-6\frac{1-\cos z}{z^{2}}.
\end{equation}
It follows from Eq. (46) that $S_{0}(z)$ is an even
function.
Using this property, we present Eq. (44) in the form
\begin{equation}
S=\frac{\pi^{2}}{Q^{2}}
\int_{0}^{1}\frac{1-x^{2}}{x^{2}} dx
\int_{0}^{\infty} U_{\ast}(k)\alpha(kQx) k^{2} dk.
\end{equation}
Formula (47) determines the denominator in Eq. (42)
for an arbitrary isotropic potential $V({\bf r})$.
For the function $U_{{\rm D}\ast}(k)$ given by Eq. (9),
this equality reads (Appendix D)
\begin{equation}
S=\frac{\pi^{2}k_{\ast}^{5}}{4}
A_{0}\Bigl (\frac{\tilde{Q}}{\delta}\Bigr ),
\end{equation}
where
\begin{equation}
A_{0}(x)= \frac{1}{x^{5}}\int_{0}^{x} (x^{2}-z^{2})^{2}
\frac{\alpha(z)}{z^{2}} dz.
\end{equation}
The function $A_{0}(x)$ is plotted in Figure 1.
This function is even,
it is negative for any $x\in (-\infty,\infty)$,
and it monotonically increases with $|x|$.
The limits of the function $A_{0}(x)$
are given by (Appendix D)
\begin{equation}
\lim_{x\to 0}A_{0}(x)=-\frac{2}{45},
\qquad
\lim_{x\to \infty} A_{0}(x)=0.
\end{equation}
It follows from Eqs. (42) and (48) that
\begin{equation}
G({\bf Q})=C\exp \Bigl [-\Bigl (\frac{3\tilde{Q}^{2}}{2}
+\frac{\varepsilon}{3\pi} A(\frac{\tilde{Q}}{\delta})
\Bigr )\Bigr ]\Bigl [1+\frac{\varepsilon}{36\pi\delta^{2}}
A_{0} (\frac{\tilde{Q}}{\delta})\Bigr ]^{-1} .
\end{equation}
Equation (51) provides an analytical expression (up
to the normalization constant $C$) for the Green
function of a flexible chain with excluded-volume
interactions.
The only assumption employed in the derivation of
this formula is the smallness of
$\varepsilon$ compared with unity (no limitations
on $\delta$ were imposed).
Given a coefficient $\delta$, Eq. (51)
is valid provided that the expression in the
last square brackets does not vanish.

Our purpose now is to determine the normalization constant $C$
and the mean square end-to-end distance $B$.
We begin with the case of ``weak" interactions
when the dimensionless parameter $\varepsilon \delta^{-2}$
(and, as a consequence, $\varepsilon$) is small
compared with unity.

\section{Weak segment interactions}

To determine the normalization constant $C$, we
substitute expression (51) into Eq. (23) and obtain
\[
4\pi b^{3} C\int_{0}^{\infty}
\exp \Bigl [-\Bigl (\frac{3\tilde{Q}^{2}}{2}
+\frac{\varepsilon}{3\pi} A(\frac{\tilde{Q}}{\delta})
\Bigr )\Bigr ]\Bigl [1
+\frac{\varepsilon}{36\pi\delta^{2}}A_{0}
(\frac{\tilde{Q}}{\delta})\Bigr ]^{-1} \tilde{Q}^{2}
d\tilde{Q}=1.
\]
Expanding the function under the integral into the
Taylor series and neglecting terms beyond the first
order of smallness with respect to $\varepsilon$
and $\varepsilon \delta^{-2}$, we find that
\[
4\pi b^{3} C\int_{0}^{\infty}
\exp \Bigl (-\frac{3\tilde{Q}^{2}}{2}\Bigr )
\Bigl \{ 1-
\frac{\varepsilon}{3\pi} \Bigl [
A(\frac{\tilde{Q}}{\delta})
+\frac{1}{12\delta^{2}}A_{0}
(\frac{\tilde{Q}}{\delta})\Bigr ]\Bigr \}
\tilde{Q}^{2} d\tilde{Q}=1.
\]
It follows from this equality that
with the required level of accuracy,
\begin{equation}
C=\Bigl (\frac{3}{2\pi b^{2}}\Bigr )^{\frac{3}{2}}
\Bigl [ 1+\varepsilon \sqrt{\frac{6}{\pi^{3}}}
\Bigl ( C_{1}+\frac{1}{12\delta^{2}}C_{2}\Bigr )\Bigr ] ,
\end{equation}
where $C_{1}$ is given by Eq. (25), and
\begin{equation}
C_{2}=\int_{0}^{\infty} A_{0} (\frac{\tilde{Q}}{\delta})
\exp \Bigl (-\frac{3\tilde{Q}^{2}}{2}\Bigr )
\tilde{Q}^{2} d\tilde{Q}.
\end{equation}
The leading term in the expression for $C_{2}$ reads
(see Appendix E for detail)
\begin{equation}
C_{2}=-\frac{4}{3}\sqrt{\frac{2\pi}{3}}\delta^{2}.
\end{equation}
Substituting expressions (26) and (54)
into Eq. (52) and neglecting terms beyond the first
order of smallness, we arrive at the formula
\begin{equation}
C=\Bigl (\frac{3}{2\pi b^{2}}\Bigr )^{\frac{3}{2}}
\Bigl ( 1-\frac{2 \varepsilon}{9\pi} \Bigr ).
\end{equation}
To calculate the mean square end-to-end distance $B$,
we substitute expression (51) into Eq. (27),
set $\tilde{Q}=Q/b$, disregard terms beyond the
first order of smallness with respect to
$\varepsilon$ and $\varepsilon \delta^{-2}$,
and find that
\[
B^{2}=4\pi b^{5}C \int_{0}^{\infty}
\exp \Bigl (-\frac{3\tilde{Q}^{2}}{2}\Bigr )
\Bigl \{ 1-\frac{\varepsilon}{3\pi}
\Bigl [ A(\frac{\tilde{Q}}{\delta})
+\frac{1}{12\delta^{2}} A_{0}(\frac{\tilde{Q}}{\delta})
\Bigr ]\Bigr \} \tilde{Q}^{4} d\tilde{Q}.
\]
It follows from this equality that
\begin{equation}
\Bigl (\frac{B}{b}\Bigr)^{2}
=C \Bigl (\frac{2\pi b^{2}}{3}\Bigr )^{\frac{3}{2}}
\Bigl [1-\varepsilon \sqrt{\frac{6}{\pi^{3}}}
\Bigl (B_{1}+\frac{1}{12\delta^{2}}B_{2}\Bigr )\Bigr ],
\end{equation}
where $B_{1}$ is given by Eq. (29), and
\begin{equation}
B_{2} = \int_{0}^{\infty}
A_{0}(\frac{\tilde{Q}}{\delta})
\exp \Bigl (-\frac{3\tilde{Q}^{2}}{2}\Bigr )
\tilde{Q}^{4} d\tilde{Q}.
\end{equation}
The leading term in the expression for $B_{2}$ reads
(Appendix E)
\begin{equation}
B_{2}=-\frac{4}{9}\sqrt{\frac{2\pi}{3}}\delta^{2} .
\end{equation}
Substituting expressions (30), (55) and (58) into
Eq. (56) and neglecting terms beyond the first order
of smallness, we find that
\begin{equation}
\Bigl (\frac{B}{b}\Bigr)^{2}
=C \Bigl (\frac{2\pi b^{2}}{3}\Bigr )^{\frac{3}{2}}
\Bigl (1+\frac{2\varepsilon}{27\pi}\Bigr )
=1-\frac{4\epsilon}{27\pi}.
\end{equation}
With reference to the conventional standpoint
[the positive contribution of $\bar{V}$ into
the Hamiltonian $H$ in Eq. (3)]
Eq. (59) contradicts our intuition:
repulsive interactions between segments
result in a decrease in the end-to-end distance.
Formula (59) contradicts also the mean-field Eq. (31):
the ratio $(B/b)^{2}$ is proportional to $\varepsilon$
instead of the classical scaling $\varepsilon\delta$.

The latter discrepancy may be explained if we
recall that excluded-volume interactions
do not permit different segments of a flexible chain
to occupy the same positions, which means that
their effect is substantial only for ``curved"
configurations of a chain, whereas the mean-field
approach is confined to ``straight" configurations.
This implies that the mean-field technique is
inapplicable to the analysis of a flexible chain
with long-range interactions between segments,
in agreement with the conclusion derived about 50
years ago \cite{Gri53} (based on different arguments).

Equation (59) leads to a physically plausible result
(an increase in the average end-to-end distance
with intensity of excluded-volume interactions)
when $\varepsilon$ is negative.
The latter is tantamount to the negativity of
contribution of $\bar{V}$ into the Hamiltonian
$H$ in Eq. (3).
To demonstrate the correctness of this assertion,
it is instructive to compare the values of $H$
on two paths: (P1) a straight line (11) that connects
the end-points, and (P2) a sinusoidal path (32) and
(34) with the only non-zero term corresponding
to a fixed $m\geq 1$.
According to the physical meaning of
excluded-volume interactions,
the energy $H_{2}$ on path (P2) should exceed
the energy $H_{1}$ on the straight path (P1)
as ${\bf Q}\to {\bf 0}$, i.e., when the chain
is superposed on itself several times.
On the other hand, it follows from Eq. (35) that
\begin{equation}
H_{2}=H_{1}+\biggl [ \frac{3\pi^{2}k_{\rm B}T}{4b^{2}}
-\frac{{v}_{0}}{(4\pi)^{2}}\int_{0}^{\infty}
\hat{U}_{{\rm D}\ast}(k)k^{4}dk\int_{-1}^{1}(1-x^{2})
p_{mm}(kQx)dx \biggr ](X_{m}^{2}+Y_{m}^{2}),
\end{equation}
where $p_{mm}$ is given by Eq. (C-29).
The integral in Eq. (60) can be calculated explicitly,
but this is not necessary for our purpose, because
Eq. (C-29) implies that in the limit of at small $Q$,
the function $p_{mm}$ reads
\[
p_{mm}(kQx)\approx L^{2}.
\]
Combining this estimate with Eq. (60), we see that
Eq. (3) with the positive contribution of ${\bar V}$
results in $H_{2}<H_{1}$ for sufficiently large
$v_{0}$, which contradicts the definition of
excluded-volume interactions.

Based on this analysis, we conclude that the potential
of intra-chain interactions should be included into
Eq. (3) with the negative sign, and, as a consequence,
the distribution function of end-to-end vectors (51)
should read
\begin{equation}
G({\bf Q})=C\exp \Bigl [-\frac{3\tilde{Q}^{2}}{2}
+\frac{2}{3}\mu\delta^{2} A(\frac{\tilde{Q}}{\delta})
\Bigr ]\Bigl [1+\mu A_{1} (\frac{\tilde{Q}}{\delta})
\Bigr ]^{-1} ,
\end{equation}
where we introduce the notation
\begin{equation}
\mu= \frac{\varepsilon}{2\pi\delta^{2}},
\qquad
A_{1}(x)=-\frac{1}{18} A_{0}(x)
\end{equation}
and preserve the positiveness of $\varepsilon\ll 1$.
We intend now to derive explicit expressions
for the normalization constant $C$ and the mean
square end-to-end distance $B$ for an arbitrary
(non necessary small) values of the dimensionless
parameter $\mu$.
Our aim is to demonstrate that Eqs. (55) and (59)
provide the leading terms in the expressions
for $C$ and $B$ (after an appropriate corrections
of signs) when the condition
$\varepsilon\delta^{-2}\ll 1$ is violated.

\section{Moderately strong interactions}

It follows from Eqs. (23) and (61) that
\[
4\pi b^{3}C\int_{0}^{\infty} \exp
\Bigl (-\frac{3x^{2}}{2}\Bigr )
\frac{1+\frac{2}{3}\mu\delta^{2}A(\frac{x}{\delta})}
{1+\mu A_{1}(\frac{x}{\delta})}x^{2}dx =1.
\]
Bearing in mind that
\[
\frac{1+\frac{2}{3}\mu\delta^{2}A(\frac{x}{\delta})}
{1+\mu A_{1}(\frac{x}{\delta})}
=1-\mu \frac{A_{1}(\frac{x}{\delta})
-\frac{2}{3}\delta^{2} A(\frac{x}{\delta})}
{1+\mu A_{1}(\frac{x}{\delta})},
\]
we find that
\begin{equation}
C=\Bigl (\frac{3}{2\pi b^{2}}\Bigr )^{\frac{3}{2}}
\biggl [1-6\mu \sqrt{\frac{3}{2 \pi}}
\int_{0}^{\infty} \frac{A_{1}(\frac{x}{\delta})
-\frac{2}{3}\delta^{2} A(\frac{x}{\delta})}
{1+\mu A_{1}(\frac{x}{\delta})}
\Bigl (-\frac{3x^{2}}{2}\Bigr )x^{2}dx \biggr ]^{-1}.
\end{equation}
Evaluating the integrals
\begin{equation}
\Lambda_{1}=\int_{0}^{\infty}
\frac{A_{1}(\frac{x}{\delta})}{1+\mu A_{1}(\frac{x}{\delta})}
\Bigl (-\frac{3x^{2}}{2}\Bigr )x^{2}dx,
\qquad
\Lambda_{2}=\int_{0}^{\infty}
\frac{A(\frac{x}{\delta})}{1+\mu A_{1}(\frac{x}{\delta})}
\Bigl (-\frac{3x^{2}}{2}\Bigr )x^{2}dx,
\end{equation}
and neglecting small terms, we arrive at the formula
(see Appendix F for detail)
\begin{equation}
C=\Bigl (\frac{3}{2\pi b^{2}}\Bigr )^{\frac{3}{2}}
\biggl \{1+\frac{\varepsilon}{\pi}\biggl [ \frac{2}{9}
-\Bigl (\frac{3}{2\pi}\Bigr )^{\frac{3}{2}}
\frac{\varepsilon}{\delta}
\int_{0}^{\infty} \frac{A_{1}^{2}(x)x^{2}}{1
+\mu A_{1}(x)}dx \biggr ] \biggr \}.
\end{equation}
Substituting Eq. (61) into Eq. (27), we find that
\begin{equation}
\Bigl (\frac{B}{b}\Bigr )^{2}=\Bigl (
\frac{2\pi b^{2}}{3}\Bigr )^{\frac{3}{2}}
C \biggl [1-3\sqrt{\frac{6}{\pi}} \mu
\int_{0}^{\infty} \frac{A_{1}(\frac{x}{\delta})
-\frac{2}{3}\delta^{2} A(\frac{x}{\delta})}
{1+\mu A_{1}(\frac{x}{\delta})}
\Bigl (-\frac{3x^{2}}{2}\Bigr )x^{4}dx\biggr ].
\end{equation}
The integrals
\begin{equation}
\Gamma_{1}=\int_{0}^{\infty}
\frac{A_{1}(\frac{x}{\delta})}{1+\mu A_{1}(\frac{x}{\delta})}
\Bigl (-\frac{3x^{2}}{2}\Bigr )x^{4}dx,
\qquad
\Gamma_{2}=\int_{0}^{\infty}
\frac{A(\frac{x}{\delta})}{1+\mu A_{1}(\frac{x}{\delta})}
\Bigl (-\frac{3x^{2}}{2}\Bigr )x^{4}dx
\end{equation}
are calculated in Appendix F, where it is shown
that the leading term in the expression for the
mean-square end-to-end distance $B$ is given by
\begin{equation}
\Bigl (\frac{B}{b}\Bigr )^{2}=\Bigl (
\frac{2\pi b^{2}}{3}\Bigr )^{\frac{3}{2}}
C \biggl \{1-\frac{2\varepsilon}{27 \pi}
\biggl [1 +\frac{81}{8\pi^{2}}\sqrt{\frac{3}{2\pi}}
\frac{\varepsilon^{2}}{\delta}
\int_{0}^{\infty} \frac{A_{1}^{3}(x)x^{4}}{1+\mu A_{1}(x)}dx
\biggr ]\biggr \}.
\end{equation}
Equations (65) and (68) imply that the leading term
in the expression for the mean square end-to-end
distance reads
\begin{equation}
\Bigl (\frac{B}{b}\Bigr )^{2}=1+\frac{\varepsilon}{\pi}
\biggl [ \frac{4}{27}
-\Bigl (\frac{3}{2\pi}\Bigr )^{\frac{3}{2}}
\frac{\varepsilon}{\delta}
\int_{0}^{\infty} \frac{A_{1}^{2}(x)x^{2}}{1+\mu A_{1}(x)}
dx \biggr ].
\end{equation}
Formulas (65) and (69) differ from Eqs. (55) and (59),
respectively, by the integral terms in the square
brackets only
(after correction of the sign of $\varepsilon$).
Although these terms contain the pre-factor
$\varepsilon/\delta$
which may accept arbitrary values, simple algebra
demonstrates (Appendix F) that their contributions
are negligible.

\section{Concluding remarks}

The formula
\[
G({\bf Q})=\Bigl (\frac{3}{2\pi b^{2}}\Bigr )^{\frac{3}{2}}
\Bigl (1+\frac{2\varepsilon}{9\pi}\Bigr )
\exp \Bigl [-\frac{3Q^{2}}{2b^{2}}
+\frac{\varepsilon}{3\pi} A(\frac{Q}{\delta b})
\Bigr ]\Bigl [1+\frac{\varepsilon}{2\pi\delta^{2}}
A_{1} (\frac{Q}{\delta b}) \Bigr ]^{-1}
\]
has been derived for the distribution function
of end-to-end vectors for a flexible chain with
weak excluded-volume interactions.
Here $b$ is the mean square end-to-end distance for
a Gaussian chain,
$\varepsilon$ is a small parameter that describes
the intensity of segment interactions,
$\delta$ is the ratio of the average end-to-end
distance for a Gaussian chain to its segment
length,
and the functions $A(x)$ and $A_{1}(x)$ are given
by Eqs. (62), (A-8) and (D-24).
The leading term in the expression for the mean square
end-to-end distance reads
\begin{equation}
\Bigl (\frac{B}{b}\Bigr )^{2}=1+\frac{4\varepsilon}{27\pi}.
\end{equation}
Equation (70) differs from similar relations developed
in previous studies, where the ratio on the left-hand side
was found to be proportional to $\varepsilon\delta$.
A reason for this difference is that our result is grounded
on the calculation of an appropriate path integral, whereas
conventional conclusions are obtained by using
mean-field approximations.
It appears that the latter approach is inapplicable to
problems where interactions between segments located
far away (along a chain) from each other are substantial,
because the mean-field technique is confined to
``straight" paths only, while inter-chain
interactions reveal themselves mainly on ``curved"
configurations of a chain.

It has been found in the calculation of the path integral
that the sign of the excluded-volume potential should
be corrected in the conventional formula (3) for the Hamiltonian.
It has also been shown that the mean square end-to-end
distance cannot be expanded into a Taylor series
with respect to small parameters (as it is traditionally
presumed), because even sub-leading terms of the highest order
in the formula for $B^{2}$ include $\delta\ln\delta$,
see Eq. (E-11), and $\sqrt{\varepsilon}$, see Eq. (F-25).

This work focuses on the analysis of the distribution
function for a flexible chain with excluded-volume
potential (9).
However, the results can be easily extended
to an arbitrary potential of intra-chain interactions
by using Eqs. (15) and (47).
In particular, these relations allow an explicit
formula to be derived for the distribution
function of end-to-end vectors for flexible
polyelectrolyte chains.
The latter will be the subject of a subsequent
publication.

\newpage
\section*{Appendix A}
\renewcommand{\theequation}{A-\arabic{equation}}
\setcounter{equation}{0}

To transform integral (14),
we choose a spherical coordinate frame
$\{ k,\phi,\theta \}$, whose $z$ axis is
directed along the vector ${\bf Q}$, and obtain
\[
\bar{V}({\bf r}_{0})=\frac{v_{0}}{2L^{2}(2\pi)^{3}}
\int_{0}^{\infty} U_{\ast}(k) k^{2} d k
\int_{0}^{2\pi} d\phi \int_{0}^{\pi} \sin \theta d\theta
\int_{0}^{L} ds\int_{0}^{L}
\exp \Bigl (- \imath k Q \cos\theta
\frac{s-s^{\prime}}{L} \Bigr ) ds^{\prime}.
\]
Calculating the integral over $\phi$ and introducing
the new variable $x=\cos\theta$, we find that
\begin{equation}
\bar{V}({\bf r}_{0})=\frac{v_{0}}{L^{2}(2\pi)^{2}}
\int_{0}^{\infty} U_{\ast}(k) k^{2} d k
\int_{0}^{1} J(kQx) dx,
\end{equation}
where
\begin{equation}
J(z)=\int_{0}^{L} ds\int_{0}^{L}
\exp \Bigl (- \imath z
\frac{s-s^{\prime}}{L} \Bigr ) ds^{\prime}
=\int_{0}^{L} \exp \Bigl (- \frac{
\imath z s}{L} \Bigr )ds
\int_{0}^{L} \exp \Bigl (\frac{\imath z s^{\prime}}{L}
\Bigr ) ds^{\prime}.
\end{equation}
For an arbitrary non-negative $z$, we have
\begin{equation}
\int_{0}^{L} \exp \Bigl (- \frac{\imath z s}{L}
\Bigr )ds  =
\frac{L}{\imath z}\Bigl (1-\exp (-\imath z)\Bigr ),
\qquad
\int_{0}^{L} \exp \Bigl (\frac{\imath z s^{\prime}}{L}
\Bigr )ds^{\prime} =
\frac{L}{\imath z}\Bigl (\exp (\imath z)-1 \Bigr ),
\end{equation}
It follows from Eqs. (A-2) and (A-3) that
\begin{equation}
J(z)=\frac{2L^{2}}{z^{2}}(1-\cos z).
\end{equation}
Substitution of Eq. (A-4) into Eq. (A-1) implies
(after changing the order of integration) that
\begin{equation}
\bar{V}({\bf r}_{0})
= \frac{v_{0}}{2\pi^{2}Q^{2}} \int_{0}^{1}
\frac{dx}{x^{2}} \int_{0}^{\infty}
U_{\ast}(k) \Bigl (1-\cos (kQx)\Bigr )dk .
\end{equation}
Combining Eqs. (12) and (A-5), we arrive at Eq. (15).

Introducing the new variable $z=kb\tilde{Q} x$,
we find from Eq. (16) that
\[
\bar{V}({\bf r}_{0})
= \frac{v_{0}}{2\pi^{2} (b\tilde{Q})^{3}} \int_{0}^{1}
\frac{dx}{x^{3}} \int_{0}^{k_{\ast}b\tilde{Q}x}
(1-\cos z)d z.
\]
Setting $y=k_{\ast}b\tilde{Q}x$, we obtain
\[
\bar{V}({\bf r}_{0})
= \frac{v_{0}k_{\ast}^{2}}{2\pi^{2} b\tilde{Q}}
\int_{0}^{k_{\ast}b\tilde{Q}}
\frac{d y}{y^{3}} \int_{0}^{y}(1-\cos z)d z.
\]
We now change the order of integration
\begin{equation}
\bar{V}({\bf r}_{0})
= \frac{v_{0}k_{\ast}^{2}}{2\pi^{2} b\tilde{Q}}
\int_{0}^{k_{\ast}b\tilde{Q}}(1-\cos z)d z
\int_{z}^{k_{\ast}b\tilde{Q}}\frac{d y}{y^{3}},
\end{equation}
and calculate the internal integral
\[
\int_{z}^{k_{\ast}b\tilde{Q}}\frac{d y}{y^{3}}
=\frac{(k_{\ast}b\tilde{Q})^{2}
-z^{2}}{2z^{2}(k_{\ast}b\tilde{Q})^{2}}.
\]
Substitution of this expression into Eq. (A-6)
implies Eqs. (17) and (18).

To determine the limits of the function $A(x)$,
we transform Eq. (18) as follows:
\begin{equation}
A(x)=\frac{1}{x} \int_{0}^{x} \frac{1-\cos z}{z^{2}}dz
-\frac{1}{x^{3}}\int_{0}^{x} (1-\cos z) dz.
\end{equation}
The first integral is simplified by integration by parts,
\[
\int_{0}^{x} \frac{1-\cos z}{z^{2}} dz
= \frac{x-\sin x}{x^{2}}
+2\int_{0}^{x} \frac{z-\sin z}{z^{3}}dz.
\]
Calculation of the second integral in Eq. (A-7)
implies that
\[
\int_{0}^{x} (1-\cos z)dz=x-\sin x.
\]
Substitution of these expressions into Eq. (A-7)
yields
\begin{equation}
A(x)=\frac{2}{x} \int_{0}^{x} a(z)dz,
\qquad
a(z) =\frac{z-\sin z}{z^{3}}.
\end{equation}
To find limits of the function $A(x)$ as $x\to 0$
and $x\to \infty$, we apply L'Hospital's rule
\begin{equation}
\lim_{x} A(x)
= \frac{2}{3}\lim_{x}\frac{1-\cos x}{x^{2}}.
\end{equation}
Equation (A-9) implies that
\[
\lim_{x\to 0} A(x)=\frac{1}{3} \lim_{x\to 0}
\frac{\sin x}{x}=\frac{1}{3},
\]
which coincides with the first equality in Eq. (19).
The other equality follows from Eq. (A-9).

\section*{Appendix B}
\renewcommand{\theequation}{B-\arabic{equation}}
\setcounter{equation}{0}

Equation (24) follows from Eq. (23) and the formula
\begin{equation}
\int_{0}^{\infty} \exp \Bigl (-\frac{3\tilde{Q}^{2}}{2}\Bigr )
\tilde{Q}^{2} d\tilde{Q}=\frac{1}{6} \sqrt{\frac{2\pi}{3}}.
\end{equation}
To determine the constant $C_{1}$, we substitute
Eq. (A-8) into Eq. (25),
introduce the new variable $x=\tilde{Q}/\delta$,
change the order of integration, and find that
\[
C_{1}=2 \delta^{3} \int_{0}^{\infty} a(z) dz
\int_{z}^{\infty}
\exp \Bigl (-\frac{3}{2} \delta^{2}x^{2} \Bigr )
x dx.
\]
Calculating the internal integral,
\begin{equation}
\int_{z}^{\infty} \exp \Bigl (-\frac{3}{2}
\delta^{2}x^{2} \Bigr )x dx
=\frac{1}{3\delta^{2}}
\exp \Bigl (-\frac{3}{2} \delta^{2}z^{2} \Bigr ),
\end{equation}
and setting $x=\delta z$, we arrive at
\[
C_{1} = \frac{2}{3} \delta \int_{0}^{\infty}
\exp \Bigl (-\frac{3}{2} \delta^{2}z^{2} \Bigr )
a(z) dz
= \frac{2}{3} \int_{0}^{\infty}
a(\frac{x}{\delta})
\exp \Bigl (-\frac{3x^{2}}{2}\Bigr )dx .
\]
It follows from Eq. (A-8) that the function
$a(x)$ is even, which implies that
\begin{equation}
C_{1} = \frac{1}{3} \int_{-\infty}^{\infty}
a(\frac{x}{\delta})
\exp \Bigl (-\frac{3x^{2}}{2}\Bigr )dx .
\end{equation}
As the limit of the function $a(x)$ as $x\to 0$
is finite
and this function decreases being proportional
to $x^{-2}$ as $x\to \infty$,
the Fourier transform
\begin{equation}
\hat{a}(s)=\int_{-\infty}^{\infty} a(x)
\exp(\imath sx)dx
\end{equation}
exists, and
\begin{equation}
a(\frac{x}{\delta})=\frac{1}{2\pi}
\int_{-\infty}^{\infty} \hat{a}(s)
\exp \Bigl (-\frac{\imath sx}{\delta}\Bigr ) ds.
\end{equation}
Substituting expression (B-5) into Eq. (B-3)
and changing the order of integration, we obtain
\begin{equation}
C_{1} = \frac{1}{6\pi} \int_{-\infty}^{\infty}
\hat{a}(s) ds\int_{-\infty}^{\infty}
\exp \Bigl [-\Bigl ( \frac{3x^{2}}{2}+
\frac{\imath sx}{\delta} \Bigr )\Bigr ]
dx .
\end{equation}
To calculate the internal integral, we set $y=x\sqrt{3}$
and find that
\begin{equation}
\int_{-\infty}^{\infty}
\exp \Bigl [-\Bigl ( \frac{3x^{2}}{2}+
\frac{\imath sx}{\delta} \Bigr )\Bigr ]dx
=\sqrt{\frac{2\pi}{3}}\exp
\Bigl (-\frac{s^{2}}{6\delta^{2}} \Bigr ).
\end{equation}
Substituting expression (B-7) into Eq. (B-6)
and introducing the new variable $x=3/(\delta\sqrt{3})$,
we obtain
\begin{equation}
C_{1} = \frac{\delta}{3\sqrt{2\pi}} \int_{-\infty}^{\infty}
\hat{a}(\delta x\sqrt{3}) \exp \Bigl (-\frac{x^{2}}{2}
\Bigr ) dx.
\end{equation}
If the function $\hat{a}(s)$ have had finite derivatives
at $s=0$, the integral were evaluated
up to an arbitrary level of accuracy with respect
to $\delta$ by the stationary phase method.
As this is not the case, we present Eq. (B-8)
in the form
\[
C_{1} = \frac{\delta}{3\sqrt{2\pi}} \Bigl [
\int_{-\infty}^{\infty} \hat{a}(0) \exp
\Bigl (-\frac{x^{2}}{2} \Bigr ) dx
+\int_{-\infty}^{\infty} \Bigl ( \hat{a}(\delta x\sqrt{3})
-\hat{a}(0)\Bigr ) \exp
\Bigl (-\frac{x^{2}}{2} \Bigr ) dx \Bigr ],
\]
calculate the first integral,
\begin{equation}
C_{1}=\frac{\delta}{3}\hat{a}(0)+R_{1},
\end{equation}
and evaluate the residual
\begin{equation}
R_{1}= \frac{\delta}{3\sqrt{2\pi}}
\int_{-\infty}^{\infty} \Bigl ( \hat{a}(\delta s\sqrt{3})
-\hat{a}(0)\Bigr ) \exp
\Bigl (-\frac{s^{2}}{2} \Bigr ) ds .
\end{equation}
It follows from Eq. (B-4) that
\[
\hat{a}(\delta s \sqrt{3})-\hat{a}(0)
=\int_{-\infty}^{\infty} a(x)
\Bigl [ \exp(\imath \delta sx\sqrt{3})
-1\Bigr ] dx .
\]
As the function $a(x)$ is even, this equality reads
\[
\hat{a}(\delta s \sqrt{3})-\hat{a}(0)
=\int_{-\infty}^{\infty} a(x)
\Bigl [ \cos(\delta sx\sqrt{3}) -1\Bigr ] dx
=-4 \int_{0}^{\infty} a(x) \sin^{2} \frac{\delta sx\sqrt{3}}{2}
dx.
\]
Substitution of Eq. (A-8) into this formula
results in
\[
\hat{a}(\delta s \sqrt{3})-\hat{a}(0)
=-3\delta^{2}s^{2}
\int_{0}^{\infty} \biggl (\frac{\sin
\frac{\delta sx\sqrt{3}}{2}}
{\frac{\delta sx\sqrt{3}}{2}}\biggr )^{2}
\Bigl (1-\frac{\sin x}{x}\Bigr ) dx.
\]
It follows from this equality that
\begin{eqnarray}
&& |\hat{a}(\delta s \sqrt{3})-\hat{a}(0)|
\leq 3\delta^{2}s^{2}
\int_{0}^{\infty} \biggl (\frac{\sin
\frac{\delta sx\sqrt{3}}{2}}
{\frac{\delta sx\sqrt{3}}{2}}\biggr )^{2}
\Bigl |1-\frac{\sin x}{x}\Bigr | dx
\nonumber\\
&&\leq  6 \delta^{2}s^{2}
\int_{0}^{\infty} \biggl (\frac{\sin
\frac{\delta sx\sqrt{3}}{2}}
{\frac{\delta sx\sqrt{3}}{2}}\biggr )^{2}dx
= 4 \delta |s|\sqrt{3}
\int_{0}^{\infty} \Bigl (\frac{\sin y}{y}\Bigr )^{2}dy
= 2\pi \delta |s| \sqrt{3}.
\end{eqnarray}
where we set $y=\frac{1}{2} \delta s x\sqrt{3}$
and use the identity
\begin{equation}
\int_{0}^{\infty} \Bigl (\frac{\sin y}{y}\Bigr )^{2} dy
=\frac{\pi}{2}.
\end{equation}
Combining Eqs. (B-10) and (B-11), we conclude
that $R_{1}$ is of order of $\delta^{2}$.
Neglecting terms beyond the first order of smallness
with respect to $\delta$, we find from Eqs. (B-4)
and (B-9) that
\begin{equation}
C_{1}
=\frac{2\delta}{3} \int_{0}^{\infty} a(x)dx.
\end{equation}
Integration by parts with the help of Eqs. (A-8)
and (B-12) results in
\begin{equation}
\int_{0}^{\infty} a(x)d x =\frac{\pi}{4}.
\end{equation}
Equations (B-13) and (B-14) result in Eq. (26).

Equation (28) follows from Eq. (27) and the identity
\begin{equation}
\int_{0}^{\infty} \exp \Bigl (-\frac{3\tilde{Q}^{2}}{2}
\Bigr )\tilde{Q}^{4} d\tilde{Q}=\frac{1}{3}
\sqrt{\frac{\pi}{6}} .
\end{equation}
To determine the coefficient $B_{1}$,
we substitute expression (A-8) into Eq. (29),
set $x=\tilde{Q}/\delta$, change the order
of integration, and find that
\[
B_{1} = 2 \delta^{5} \int_{0}^{\infty}
a(z)dz \int_{z}^{\infty}
\exp \Bigl (-\frac{3\delta^{2}x^{2}}{2}\Bigr ) x^{3}
dx.
\]
Bearing in mind that
\begin{equation}
\int_{z}^{\infty} \exp
\Bigl (-\frac{3\delta^{2}x^{2}}{2}\Bigr ) x^{3} dx
=\frac{1}{3\delta^{2}}\Bigl (z^{2}+\frac{2}{3\delta^{2}}
\Bigr )\exp \Bigl (-\frac{3\delta^{2}z^{2}}{2}\Bigr ),
\end{equation}
we obtain
\begin{equation}
B_{1}= \frac{4}{9}\Bigl [
\int_{0}^{\infty}a(\frac{y}{\delta})\exp
\Bigl (-\frac{3y^{2}}{2}\Bigr )d y
+\frac{3}{2} \int_{0}^{\infty}a(\frac{y}{\delta})\exp
\Bigl (-\frac{3 y^{2}}{2}\Bigr )y^{2} dy \Bigr ].
\end{equation}
The first integral is given by Eq. (B-3).
To determine the other integral, we use
Eq. (B-5) for the function $a$ and
change the order of integration,
\begin{equation}
\int_{0}^{\infty} a(\frac{y}{\delta})\exp
\Bigl (-\frac{3y^{2}}{2} \Bigr )y^{2} dy
= \frac{1}{4\pi}\int_{-\infty}^{\infty}
\hat{a}(s) ds \int_{-\infty}^{\infty}
\exp \Bigl [-\Bigl (\frac{3y^{2}}{2}
+\frac{\imath sy}{\delta} \Bigr )\Bigr ]y^{2}dy.
\end{equation}
The internal integral in Eq. (B-18) is calculated
explicitly
\begin{equation}
\int_{-\infty}^{\infty}
\exp \Bigl [-\Bigl (\frac{3y^{2}}{2}
+\frac{\imath sy}{\delta} \Bigr )\Bigr ]y^{2}dy
=\frac{1}{3}\sqrt{\frac{2\pi}{3}}
\exp \Bigl (-\frac{s^{2}}{6\delta^{2}}\Bigr )
\Bigl (1-\frac{s^{2}}{3\delta^{2}}\Bigr ).
\end{equation}
Substitution of Eqs. (B-3), (B-18) and (B-19)
into Eq. (B-17) results in
\begin{equation}
B_{1}= \frac{2}{3} \Bigl [ C_{1}+
\frac{\delta}{6\sqrt{2\pi}}
\int_{-\infty}^{\infty}
\hat{a}(\delta z \sqrt{3}) \exp \Bigl (-\frac{z^{2}}{2}
\Bigr )(1-z^{2})d z \Bigr ],
\end{equation}
where $z=s/(\delta \sqrt{3})$.
Evaluating the integral by using the same approach
that was applied to derive Eq. (B-13)
and utilizing the identity
\begin{equation}
\int_{-\infty}^{\infty} \exp \Bigl (-\frac{z^{2}}{2}
\Bigr )(1-z^{2})d z =0,
\end{equation}
we find that the order of smallness of the second term
on the right-hand side of Eq. (B-20) is higher than
that of the first term.
Equation (30) follows from Eqs. (B-13), (B-14) and (B-20).

\section*{Appendix C}
\renewcommand{\theequation}{C-\arabic{equation}}
\setcounter{equation}{0}

Substitution of expressions (32) to (34) into Eq. (1)
results in
\[
H_{0}({\bf r})
= \frac{3k_{\rm B}T}{2b_{0}}\biggl [\frac{Q^{2}}{L}
+\sum_{m,n=1}^{\infty} \frac{\pi m}{L}\frac{\pi n}{L}
(X_{m}X_{n}+Y_{m}Y_{n})\int_{0}^{L} \cos\frac{\pi m s}{L}
\cos\frac{\pi n s}{L} ds \biggr ] .
\]
Employing the orthogonality of the trigonometric
functions and Eq. (12), we arrive at
\begin{equation}
H_{0}({\bf r}) = \frac{3k_{\rm B}T}{2b^{2}}
\Bigl [Q^{2}
+\frac{\pi^{2}}{2}\sum_{m=1}^{\infty} m^{2}(X_{m}^{2}
+Y_{m}^{2}) \Bigr ].
\end{equation}
The first equality in Eq. (35) follows from Eq. (C-1).

Substituting Eq. (32) into the exponent in Eq. (13),
expanding the obtained expression into the Taylor
series, and omitting terms beyond the second order
of smallness with respect to $|{\bf R}|$,
we find that
\begin{eqnarray}
\exp \Bigl [ -\imath {\bf k}\cdot \Bigl ({\bf r}(s)
-{\bf r}(s^{\prime})\Bigr )\Bigr ]
&=& \exp \Bigl [ -\imath {\bf k}\cdot {\bf Q}
\frac{s-s^{\prime}}{L}\Bigr ]\biggl \{ 1
-\imath {\bf k}\cdot \Bigl ({\bf R}(s)
-{\bf R}(s^{\prime})\Bigr )
\nonumber\\
&& -\frac{1}{2} \Bigl [ {\bf k}\cdot\Bigl ({\bf R}(s)
-{\bf R}(s^{\prime})\Bigr )\Bigr ]^{2}\biggr \}.
\end{eqnarray}
Bearing in mind that ${\bf Q}=Q{\bf e}_{3}$,
we conclude from Eq. (34) that
\[
{\bf k}\cdot {\bf Q}=k_{3}Q,
\qquad
{\bf k}\cdot {\bf R}(s)=\sum_{m=1}^{\infty} (k_{1}X_{m}
+k_{2}Y_{m})\sin \frac{\pi m s}{L}.
\]
Substitution of these expressions into Eq. (C-2) results in
\begin{eqnarray}
&& \exp \Bigl [ -\imath {\bf k}\cdot \Bigl ({\bf r}(s)
-{\bf r}(s^{\prime})\Bigr )\Bigr ]
= \exp \Bigl ( -\imath k_{3}Q \frac{s-s^{\prime}}{L}
\Bigr )\biggl [ 1
-\imath \sum_{m=1}^{\infty} (k_{1} X_{m}+k_{2}Y_{m})
\nonumber\\
&& \times
\Bigl ( \sin \frac{\pi ms}{L}
-\sin\frac{\pi m s^{\prime}}{L}\Bigr )
-\frac{1}{2} \sum_{m,n=1}^{\infty}
(k_{1} X_{m}+k_{2}Y_{m})
\Bigl ( \sin \frac{\pi ms}{L}
-\sin\frac{\pi m s^{\prime}}{L}\Bigr )
\nonumber\\
&&\times
(k_{1} X_{n}+k_{2}Y_{n})
\Bigl ( \sin \frac{\pi n s}{L}
-\sin\frac{\pi n s^{\prime}}{L}\Bigr )\biggr ].
\end{eqnarray}
Combining Eq. (C-3) with Eqs. (13) and (14),
we find that
\begin{equation}
\bar{V}({\bf r})=\bar{V}({\bf r}_{0})
+ \frac{v_{0}}{2L^{2}(2\pi)^{3}}
\Bigl [ \bar{V}_{1}({\bf R})+\bar{V}_{2}({\bf R})\Bigr ],
\end{equation}
where
\begin{equation}
\bar{V}_{1}=\int U({\bf k}) P_{1}({\bf k})d{\bf k},
\qquad
\bar{V}_{2}=\int U({\bf k}) P_{2}({\bf k})d{\bf k}.
\end{equation}
The functions $P_{1}$ and $P_{2}$ in Eq. (C-5)
are given by
\begin{eqnarray}
P_{1} ({\bf k}) &=& -\frac{\imath}{2} \sum_{m=1}^{\infty}
(k_{1}X_{m}+k_{2}Y_{m}) p_{m}(k_{3}Q),
\nonumber\\
P_{2} ({\bf k}) &=& -\frac{1}{2} \sum_{m,n=1}^{\infty}
(k_{1}X_{m}+k_{2}Y_{m})(k_{1}X_{n}+k_{2}Y_{n}) p_{mn}(k_{3}Q),
\end{eqnarray}
where the coefficients $p_{m}$ and $p_{mn}$ read
\begin{eqnarray}
\hspace*{-10 mm}&& p_{m}(k_{3}Q) = \int_{0}^{L} ds\int_{0}^{L}
\exp \Bigl ( -\imath k_{3}Q \frac{s-s^{\prime}}{L}
\Bigr )
\Bigl ( \sin \frac{\pi ms}{L}
-\sin\frac{\pi m s^{\prime}}{L} \Bigr )ds^{\prime},
\nonumber\\
\hspace*{-10 mm}&&p_{mn} (k_{3}Q) =
\int_{0}^{L} ds\int_{0}^{L}
\exp \Bigl ( -\imath k_{3}Q \frac{s-s^{\prime}}{L}
\Bigr )
\Bigl ( \sin \frac{\pi ms}{L}
-\sin\frac{\pi m s^{\prime}}{L} \Bigr )
\Bigl ( \sin \frac{\pi n s}{L}
-\sin\frac{\pi n s^{\prime}}{L} \Bigr )ds^{\prime}.
\end{eqnarray}
In spherical coordinates $\{ k,\phi, \theta \}$, the Cartesian
components of the vector ${\bf k}$ read
\[
k_{1}=k\sin\theta\cos\phi,
\quad
k_{2}=k\sin\theta\sin\phi,
\quad
k_{3}=k\cos\theta .
\]
Substitution of these relations
into Eqs. (C-5) and (C-6) implies that
\begin{equation}
\bar{V}_{1} = -\frac{\imath}{2}\sum_{m=1}^{\infty}
\int_{0}^{\infty} U_{\ast}(k) k^{3} dk
\int_{0}^{2\pi} \Bigl (X_{m}\cos\phi+Y_{m}\sin \phi\Bigr )
d\phi \int_{0}^{\pi} p_{m}(kQ\cos\theta)
\sin^{2}\theta d\theta =0.
\end{equation}
By analogy with Eq. (C-8), we write
\begin{eqnarray*}
\bar{V}_{2} &=& -\frac{1}{2} \sum_{m,n=1}^{\infty}
\int_{0}^{\infty} U_{\ast}(k) k^{4} dk
\int_{0}^{2\pi} \Bigl [ X_{m}X_{n}\cos^{2}\phi
\nonumber\\
&& +(X_{m}Y_{n}+X_{n}Y_{m})\sin\phi\cos\phi
+Y_{m}Y_{n}\sin^{2} \phi\Bigr ] d\phi \int_{0}^{\pi}
p_{mn}(kQ\cos\theta) \sin^{3}\theta d\theta.
\end{eqnarray*}
Calculating the integral over $\phi$ and introducing the
new variable $x=\cos \theta$, we arrive at
\begin{equation}
\bar{V}_{2} = -\frac{\pi}{2}
\sum_{m,n=1}^{\infty} (X_{m}X_{n}+Y_{m}Y_{n})
\int_{0}^{\infty} U_{\ast}(k) k^{4} dk
\int_{-1}^{1} p_{mn}(kQx)(1-x^{2})dx.
\end{equation}
Substitution of expressions (C-8) and (C-9) into
Eq. (C-4) implies that
\begin{equation}
\bar{V}({\bf r})=\bar{V}({\bf r}_{0})
-\frac{v_{0}}{32\pi^{2}L^{2}}
\sum_{m,n=1}^{\infty} (X_{m}X_{n}
+Y_{m}Y_{n})\int_{0}^{\infty} U_{\ast}(k) k^{4} dk
\int_{-1}^{1} p_{mn}(kQx)(1-x^{2})dx .
\end{equation}
Our aim now is to determine the coefficients $p_{mn}$.
It follows from Eq. (C-7) that
\begin{equation}
p_{mn}=p_{mn}^{(1)}-p_{mn}^{(2)}-p_{nm}^{(2)}+p_{mn}^{(3)},
\end{equation}
where
\begin{eqnarray}
p_{mn}^{(1)} &=& \int_{0}^{L} \exp \Bigl (-\imath k_{3}Q\frac{s}{L}
\Bigr )\sin \frac{\pi m s}{L}\sin \frac{\pi n s}{L} ds
\int_{0}^{L} \exp \Bigl (\imath k_{3}Q \frac{s^{\prime}}{L}
\Bigr )ds^{\prime},
\nonumber\\
p_{mn}^{(2)} &=& \int_{0}^{L} \exp \Bigl (-\imath k_{3}Q\frac{s}{L}
\Bigr )\sin \frac{\pi n s}{L}ds\int_{0}^{L} \exp \Bigl (
\imath k_{3}Q \frac{s^{\prime}}{L}\Bigr )
\sin \frac{\pi m s^{\prime}}{L} ds^{\prime},
\nonumber\\
p_{mn}^{(3)}  &=& \int_{0}^{L} \exp \Bigl (-\imath k_{3}Q\frac{s}{L}
\Bigr )ds\int_{0}^{L} \exp \Bigl ( \imath k_{3}Q \frac{s^{\prime}}{L}\Bigr )
\sin \frac{\pi m s^{\prime}}{L} \sin \frac{\pi n s^{\prime}}{L} ds^{\prime} .
\end{eqnarray}
We begin with the quantity
\begin{equation}
B_{mn}=\int_{0}^{L} \exp \Bigl (-\imath k_{3}Q\frac{s}{L}
\Bigr )\sin \frac{\pi m s}{L}\sin \frac{\pi n s}{L} ds
=\frac{1}{2} (B_{mn}^{(1)}-B_{mn}^{(2)}),
\end{equation}
where
\begin{eqnarray}
B_{mn}^{(1)} &=& \int_{0}^{L}
\exp \Bigl (-\imath k_{3}Q\frac{s}{L}
\Bigr )\cos \frac{\pi(m-n)s}{L} ds,
\nonumber\\
B_{mn}^{(2)} &=& \int_{0}^{L}
\exp \Bigl (-\imath k_{3}Q\frac{s}{L}
\Bigr )\cos \frac{\pi(m+n)s}{L} ds.
\end{eqnarray}
Using the formula
$\cos \frac{\pi l s}{L}=\frac{1}{2} \Bigl [
\exp \Bigl (\frac{\imath \pi l s}{L}\Bigr )
+\exp \Bigl (-\frac{\imath \pi l s}{L}\Bigr )
\Bigr ]$
and calculating the integral in Eq. (C-14),
we obtain
\begin{eqnarray*}
B_{mn}^{(1)} &=& \frac{1}{2} \int_{0}^{L}
\biggl [ \exp \Bigl ( -\imath (k_{3}Q-\pi(m-n))
\frac{s}{L}\Bigr )
+\exp \Bigl ( -\imath (k_{3}Q+\pi(m-n))
\frac{s}{L}\Bigr )\biggr ]ds
\nonumber\\
&=& -\frac{L}{2\imath} \biggl [
\frac{1}{k_{3}Q-\pi(m-n)}
\Bigl (\exp\Bigl (-\imath (k_{3}Q-\pi (m-n))\Bigr )
-1\Bigr )
\nonumber\\
&& +\frac{1}{k_{3}Q+\pi(m-n)}
\Bigl (\exp\Bigl (-\imath (k_{3}Q+\pi (m-n))\Bigr )
-1\Bigr ) \biggr ] .
\end{eqnarray*}
Taking into account that
\begin{equation}
\exp (-\imath x)-1
=-2\imath \sin \frac{x}{2} \exp(-\frac{\imath x}{2}),
\end{equation}
we find that
\begin{eqnarray*}
B_{mn}^{(1)} &=& L \exp \Bigl (-\frac{\imath}{2} k_{3}Q\Bigr )
\biggl [ \exp \Bigl (\frac{\pi}{2} \imath (m-n)\Bigr )
F(k_{3}Q-\pi(m-n))
\nonumber\\
&& +\exp \Bigl (-\frac{\pi}{2} \imath (m-n)\Bigr )
F(k_{3}Q+\pi(m-n))\biggr ],
\end{eqnarray*}
where the function $F(z)$ is determined by Eq. (37).
Similar transformations result in
\begin{eqnarray*}
B_{mn}^{(2)} &=& L \exp \Bigl (-\frac{\imath}{2} k_{3}Q\Bigr )
\biggl [ \exp \Bigl (\frac{\pi}{2} \imath (m+n)\Bigr )
F(k_{3}Q-\pi(m+n))
\nonumber\\
&& +\exp \Bigl (-\frac{\pi}{2} \imath (m+n)\Bigr )
F(k_{3}Q+\pi(m+n))\biggr ].
\end{eqnarray*}
Substitution of these expressions into Eq. (C-13)
implies that
\begin{eqnarray}
B_{mn} &=& \frac{L}{2}\exp \Bigl (-\frac{\imath}{2} k_{3}Q\Bigr )
\biggl \{
\biggl [ \exp \Bigl (\frac{\pi}{2} \imath (m-n)\Bigr )
F(k_{3}Q-\pi(m-n))
\nonumber\\
&& +\exp \Bigl (-\frac{\pi}{2} \imath (m-n)\Bigr )
F(k_{3}Q+\pi(m-n))\biggr ]
\nonumber\\
&& -\biggl [ \exp \Bigl (\frac{\pi}{2} \imath (m+n)\Bigr )
F(k_{3}Q-\pi(m+n))
\nonumber\\
&& +\exp \Bigl (-\frac{\pi}{2} \imath (m+n)\Bigr )
F(k_{3}Q+\pi(m+n))\biggr ] \biggr \}.
\end{eqnarray}
It follows from Eqs. (37), (A-3) and (C-15) that
\begin{equation}
\int_{0}^{L}\exp \Bigl (\imath k_{3}Q \frac{s^{\prime}}{L}
\Bigr )ds^{\prime}
= 2L\exp (\frac{\imath}{2} k_{3}Q)F(k_{3}Q).
\end{equation}
Substitution of Eqs. (C-16) and (C-17) into Eq. (C-12)
results in
\begin{eqnarray*}
p_{mn}^{(1)} &=& L^{2}F(k_{3}Q)
\biggl \{
\biggl [ \exp \Bigl (\frac{\pi}{2} \imath (m-n)\Bigr )
F(k_{3}Q-\pi(m-n))
\nonumber\\
&& +\exp \Bigl (-\frac{\pi}{2} \imath (m-n)\Bigr )
F(k_{3}Q+\pi(m-n))\biggr ]
\nonumber\\
&& -\biggl [ \exp \Bigl (\frac{\pi}{2} \imath (m+n)\Bigr )
F(k_{3}Q-\pi(m+n))
\nonumber\\
&& +\exp \Bigl (-\frac{\pi}{2} \imath (m+n)\Bigr )
F(k_{3}Q+\pi(m+n))\biggr ] \biggr \}.
\end{eqnarray*}
Replacing the exponents by trigonometric functions,
we find that
\begin{eqnarray}
p_{mn}^{(1)} &=& L^{2}F(k_{3}Q)
\biggl \langle \biggl \{\cos\frac{\pi(m-n)}{2}
\Bigl [ F(k_{3}Q+\pi(m-n))+F(k_{3}Q-\pi(m-n))\Bigr ]
\nonumber\\
&& -\cos\frac{\pi(m+n)}{2}
\Bigl [ F(k_{3}Q+\pi(m+n))+F(k_{3}Q-\pi(m+n))
\Bigr ]\biggr \}
\nonumber\\
&&+\imath \biggl \{ -\sin\frac{\pi(m-n)}{2}
\Bigl [ F(k_{3}Q+\pi(m-n))-F(k_{3}Q-\pi(m-n))\Bigr ]
\nonumber\\
&& +\sin\frac{\pi(m+n)}{2}
\Bigl [ F(k_{3}Q+\pi(m+n))-F(k_{3}Q-\pi(m+n))\Bigr ]
\biggr \}\biggr \rangle .
\end{eqnarray}
We proceed with transformation of the quantity
\begin{equation}
C_{mn}=\int_{0}^{L} \exp \Bigl (
\imath k_{3}Q\frac{s^{\prime}}{L}
\Bigr )\sin \frac{\pi m s^{\prime}}{L}
\sin \frac{\pi n s^{\prime}}{L} ds^{\prime}.
\end{equation}
Comparison of expressions (C-13) and (C-19) implies
that $C_{mn}(k_{3})$ coincides with $B_{mn}(-k_{3})$.
According to Eq. (C-16), this means that
\begin{eqnarray*}
C_{mn} &=& \frac{L}{2}\exp \Bigl (
\frac{\imath}{2} k_{3}Q\Bigr )
\biggl \{ \biggl [ \exp \Bigl (\frac{\pi}{2}
\imath (m-n)\Bigr ) F(-k_{3}Q-\pi(m-n))
\nonumber\\
&& +\exp \Bigl (-\frac{\pi}{2} \imath (m-n)\Bigr )
F(-k_{3}Q+\pi(m-n))\biggr ]
\nonumber\\
&& -\biggl [ \exp \Bigl (\frac{\pi}{2}
\imath (m+n)\Bigr ) F(-k_{3}Q-\pi(m+n))
\nonumber\\
&& +\exp \Bigl (-\frac{\pi}{2} \imath (m+n)\Bigr )
F(-k_{3}Q+\pi(m+n))\biggr ] \biggr \}.
\end{eqnarray*}
Bearing in mind that the function $F(z)$ is even,
we find that
\begin{eqnarray}
C_{mn} &=& \frac{L}{2}\exp \Bigl (\frac{\imath}{2}
k_{3}Q\Bigr ) \biggl \{
\biggl [ \exp \Bigl (\frac{\pi}{2} \imath (m-n)\Bigr )
F(k_{3}Q+\pi(m-n))
\nonumber\\
&& +\exp \Bigl (-\frac{\pi}{2} \imath (m-n)\Bigr )
F(k_{3}Q-\pi(m-n))\biggr ]
\nonumber\\
&& -\biggl [ \exp \Bigl (\frac{\pi}{2} \imath (m+n)\Bigr )
F(k_{3}Q+\pi(m+n))
\nonumber\\
&& +\exp \Bigl (-\frac{\pi}{2} \imath (m+n)\Bigr )
F(k_{3}Q-\pi(m+n))\biggr ] \biggr \}.
\end{eqnarray}
It follows from Eq. (C-17) that
\begin{equation}
\int_{0}^{L} \exp \Bigl (-\imath k_{3}Q\frac{s}{L} \Bigr )ds
=2L \exp (-\frac{\imath}{2} k_{3}Q)F(k_{3}Q).
\end{equation}
Substitution of expressions (C-20) and (C-21)
into Eq. (C-12) results in
\begin{eqnarray*}
p_{mn}^{(3)} &=& L^{2}F(k_{3}Q)
\biggl \{
\biggl [ \exp \Bigl (\frac{\pi}{2} \imath (m-n)\Bigr )
F(k_{3}Q+\pi(m-n))
\nonumber\\
&& +\exp \Bigl (-\frac{\pi}{2} \imath (m-n)\Bigr )
F(k_{3}Q-\pi(m-n))\biggr ]
\nonumber\\
&& -\biggl [ \exp \Bigl (\frac{\pi}{2} \imath (m+n)\Bigr )
F(k_{3}Q+\pi(m+n))
\nonumber\\
&& +\exp \Bigl (-\frac{\pi}{2} \imath (m+n)\Bigr )
F(k_{3}Q-\pi(m+n))\biggr ] \biggr \}.
\end{eqnarray*}
In the trigonometric form, this equality reads
\begin{eqnarray}
p_{mn}^{(3)} &=& L^{2}F(k_{3}Q)
\biggl \langle \biggl \{\cos\frac{\pi(m-n)}{2}
\Bigl [ F(k_{3}Q+\pi(m-n))+F(k_{3}Q-\pi(m-n))\Bigr ]
\nonumber\\
&& -\cos\frac{\pi(m+n)}{2}
\Bigl [ F(k_{3}Q+\pi(m+n))+F(k_{3}Q-\pi(m+n))\Bigr ]\biggr \}
\nonumber\\
&&+\imath \biggl \{ \sin\frac{\pi(m-n)}{2}
\Bigl [ F(k_{3}Q+\pi(m-n))-F(k_{3}Q-\pi(m-n))\Bigr ]
\nonumber\\
&& -\sin\frac{\pi(m+n)}{2}
\Bigl [ F(k_{3}Q+\pi(m+n))-F(k_{3}Q-\pi(m+n))\Bigr ]
\biggr \}\biggr \rangle .
\end{eqnarray}
Combining Eqs. (C-18) and (C-22), we find that
\begin{eqnarray}
p_{mn}^{(1)}+p_{mn}^{(3)} &=& L^{2}F(k_{3}Q)
\biggl \{\cos\frac{\pi(m-n)}{2}
\Bigl [ F(k_{3}Q+\pi(m-n))+F(k_{3}Q-\pi(m-n))\Bigr ]
\nonumber\\
&& -\cos\frac{\pi(m+n)}{2}
\Bigl [ F(k_{3}Q+\pi(m+n))+F(k_{3}Q-\pi(m+n))
\Bigr ]\biggr \}.
\end{eqnarray}
We now calculate the integrals
\begin{equation}
D_{m}^{\pm }=\int_{0}^{L} \exp
\Bigl (\pm \imath k_{3}Q\frac{s}{L}
\Bigr )\sin \frac{\pi m s}{L}ds.
\end{equation}
Bearing in mind that
$\sin \frac{\pi m s}{L}=\frac{1}{2\imath} \Bigl [
\exp \Bigl (\frac{\imath \pi m s}{L}\Bigr )
-\exp \Bigl (-\frac{\imath \pi m s}{L}\Bigr )\Bigr ]$,
we find that
\begin{eqnarray*}
D_{m}^{+}
&=& \frac{1}{2\imath} \biggl [ \int_{0}^{L}
\exp \Bigl (\imath (k_{3}Q+\pi m)\frac{s}{L}\Bigr ) ds
-\int_{0}^{L} \exp \Bigl (\imath (k_{3}Q-\pi m)\frac{s}{L}
\Bigr ) ds \biggr ]
\nonumber\\
&=& \frac{L}{2} \biggl [ \frac{1}{k_{3}Q-\pi m}
\Bigl (\exp (\imath (k_{3}Q-\pi m)) -1 \Bigr )
\nonumber\\
&& -\frac{1}{k_{3}Q+\pi m}
\Bigl (\exp (\imath (k_{3}Q+\pi m))-1 \Bigr )\biggr ] .
\end{eqnarray*}
It follows from this equality and Eq. (C-15) that
\[
D_{m}^{+} = L \imath \exp \Bigl (\frac{\imath}{2}k_{3}Q\Bigr )
\biggl [\exp \Bigl (-\frac{\pi}{2}\imath m\Bigr )F(k_{3}Q-\pi m)
-\exp \Bigl (\frac{\pi}{2}\imath m\Bigr )F(k_{3}Q+\pi m) \biggr ].
\]
Similarly, we find that
\[
D_{m}^{-} = L \imath \exp \Bigl (-\frac{\imath}{2}k_{3}Q\Bigr )
\biggl [\exp \Bigl (-\frac{\pi}{2}\imath m\Bigr )F(k_{3}Q+\pi m)
-\exp \Bigl (\frac{\pi}{2}\imath m\Bigr )F(k_{3}Q-\pi m) \biggr ].
\]
Substitution of these expressions into Eq. (C-12)
implies that
\begin{eqnarray*}
p_{mn}^{(2)} &=& L^{2}\biggl [
\exp \Bigl (\frac{\pi}{2}\imath m\Bigr )F(k_{3}Q+\pi m)
-\exp \Bigl (-\frac{\pi}{2}\imath m\Bigr )F(k_{3}Q-\pi m)\biggr ]
\nonumber\\
&&\times
\biggl [\exp \Bigl (-\frac{\pi}{2}\imath n\Bigr )F(k_{3}Q+\pi n)
-\exp \Bigl (\frac{\pi}{2}\imath n\Bigr )F(k_{3}Q-\pi n) \biggr ].
\end{eqnarray*}
Taking into account that
$\exp \Bigl (\frac{\pi}{2}\imath n\Bigr )
=\cos\frac{\pi n}{2}+\imath \sin\frac{\pi n}{2}$,
we arrive at the formula
\begin{eqnarray*}
p_{mn}^{(2)} &=& L^{2}\Bigl [ \cos\frac{\pi m}{2}
\Bigl ( F(k_{3}Q+\pi m)- F(k_{3}Q-\pi m)\Bigr )
\nonumber\\
&& +\imath\sin\frac{\pi m}{2}
\Bigl ( F(k_{3}Q+\pi m)+ F(k_{3}Q-\pi m)\Bigr )\Bigr ]
\nonumber\\
&&\times
\Bigl [ \cos\frac{\pi n}{2}
\Bigl ( F(k_{3}Q+\pi n)- F(k_{3}Q-\pi n)\Bigr )
\nonumber\\
&& -\imath\sin\frac{\pi n}{2}
\Bigl ( F(k_{3}Q+\pi n)+ F(k_{3}Q-\pi n)\Bigr )\Bigr ].
\end{eqnarray*}
It follows from this equality that
\begin{equation}
p_{mn}^{(2)}=L^{2}(r_{mn}+\imath s_{mn}),
\end{equation}
where
\begin{eqnarray*}
r_{mn} &=& \cos\frac{\pi m}{2}\cos\frac{\pi n}{2}
\Bigl [ F(k_{3}Q+\pi m)- F(k_{3}Q-\pi m)\Bigr ]
\Bigl [ F(k_{3}Q+\pi n)- F(k_{3}Q-\pi n)\Bigr ]
\nonumber\\
&& +\sin\frac{\pi m}{2}\sin\frac{\pi n}{2}
\Bigl [ F(k_{3}Q+\pi m)+ F(k_{3}Q-\pi m) \Bigr ]
\Bigl [ F(k_{3}Q+\pi n)+F(k_{3}Q-\pi n) \Bigr ],
\nonumber\\
s_{mn} &=& \sin\frac{\pi m}{2}\cos\frac{\pi n}{2}
\Bigl [ F(k_{3}Q+\pi m)+ F(k_{3}Q-\pi m)\Bigr ]
\Bigl [ F(k_{3}Q+\pi n)- F(k_{3}Q-\pi n)\Bigr ]
\nonumber\\
&& -\cos\frac{\pi m}{2}\sin\frac{\pi n}{2}
\Bigl [ F(k_{3}Q+\pi m)- F(k_{3}Q-\pi m) \Bigr ]
\Bigl [ F(k_{3}Q+\pi n)+F(k_{3}Q-\pi n) \Bigr ].
\end{eqnarray*}
As $r_{nm}=r_{mn}$ and $s_{nm}=-s_{mn}$,
we find from Eq. (C-25) that
\begin{eqnarray}
p_{mn}^{(2)}+p_{nm}^{(2)} &=& 2L^{2} \Bigl \{
\cos\frac{\pi m}{2}\cos\frac{\pi n}{2}
\Bigl [ F(k_{3}Q+\pi m)- F(k_{3}Q-\pi m)\Bigr ]
\nonumber\\
&&\times
\Bigl [ F(k_{3}Q+\pi n)- F(k_{3}Q-\pi n)\Bigr ]
\nonumber\\
&& +\sin\frac{\pi m}{2}\sin\frac{\pi n}{2}
\Bigl [ F(k_{3}Q+\pi m)+ F(k_{3}Q-\pi m) \Bigr ]
\nonumber\\
&&\times
\Bigl [ F(k_{3}Q+\pi n)+F(k_{3}Q-\pi n) \Bigr ]
\Bigr \}.
\end{eqnarray}
Substitution of expressions (C-23) and (C-26)
into Eq. (C-11) results in
\begin{eqnarray}
p_{mn} &=& 2L^{2}
\Bigl \{ F(k_{3}Q)\Bigl [ \cos\frac{\pi(m-n)}{2}
\Bigl ( F(k_{3}Q+\pi(m-n))+F(k_{3}Q-\pi (m-n))\Bigr )
\nonumber\\
&&-\cos\frac{\pi(m+n)}{2}
\Bigl ( F(k_{3}Q+\pi(m+n))+F(k_{3}Q-\pi (m+n))\Bigr )\Bigr ]
\nonumber\\
&&- \cos\frac{\pi m}{2}\cos\frac{\pi n}{2}
\Bigl [ F(k_{3}Q+\pi m)- F(k_{3}Q-\pi m)\Bigr ]
\nonumber\\
&&\times
\Bigl [ F(k_{3}Q+\pi n)- F(k_{3}Q-\pi n)\Bigr ]
-\sin\frac{\pi m}{2}\sin\frac{\pi n}{2}
\Bigl [ F(k_{3}Q+\pi m)
\nonumber\\
&&+ F(k_{3}Q-\pi m) \Bigr ]
\Bigl [ F(k_{3}Q+\pi n)+F(k_{3}Q-\pi n) \Bigr ]
\Bigr \}.
\end{eqnarray}
It follows from Eqs. (C-10) and (C-27) that the
potential $\bar{V}({\bf r})$ is given by Eq. (35),
where the coefficients
\begin{equation}
P_{mn}=\frac{1}{2L^{2}}\int_{0}^{\infty} U_{\ast}(k)k^{4}
dk \int_{-1}^{1} p_{mn}(kQx)(1-x^{2}) dx
\end{equation}
are determined by Eq. (36).

Setting $n=m$ in Eq. (C-27), we find that
\begin{eqnarray*}
p_{m m} &=& 2L^{2}
\Bigl \{ F(k_{3}Q) \Bigl [ 2 F(k_{3}Q)
-\cos (\pi m) \Bigl ( F(k_{3}Q+2\pi m)
+F(k_{3}Q-2 \pi m)\Bigr )\Bigr ]
\nonumber\\
&&- \cos^{2} \frac{\pi m}{2}
\Bigl [ F(k_{3}Q+\pi m)- F(k_{3}Q-\pi m)\Bigr ]^{2}
-\sin^{2} \frac{\pi m}{2}\Bigl [ F(k_{3}Q+\pi m)
\nonumber\\
&&+ F(k_{3}Q-\pi m) \Bigr ]^{2} \Bigr \}.
\end{eqnarray*}
Bearing in mind that
\begin{eqnarray*}
&&\sin^{2}\frac{\pi m}{2} \Bigl (
F(k_{3}Q+\pi m)+F(k_{3}Q-\pi m)\Bigr )^{2}
+\cos^{2}\frac{\pi m}{2}
\Bigl (F(k_{3}Q+\pi m)-F(k_{3}Q-\pi m)\Bigr )^{2}
\nonumber\\
&&= F^{2}(k_{3}Q+\pi m)+F^{2}(k_{3}Q-\pi m)
-2\cos(\pi m)F(k_{3}Q+\pi m)F(k_{3}Q-\pi m),
\end{eqnarray*}
we arrive at the formula
\begin{eqnarray}
p_{m m} &=& 2L^{2}
\Bigl \{ \Bigl [ 2 F^{2}(k_{3}Q)-F^{2}(k_{3}Q+\pi m)
-F^{2}(k_{3}Q-\pi m)\Bigr ]
\nonumber\\
&& -\cos (\pi m) \Bigl [ F(k_{3}Q)\Bigl ( F(k_{3}Q+2\pi m)
+F(k_{3}Q-2 \pi m)\Bigr )
\nonumber\\
&& -2 F(k_{3}Q+\pi m)F(k_{3}Q-\pi m) \Bigr ]\Bigr \}.
\end{eqnarray}

\section*{Appendix D}
\renewcommand{\theequation}{D-\arabic{equation}}
\setcounter{equation}{0}

It follows from Eqs. (43), (C-28) and (C-29) that
\begin{eqnarray*}
S &=& \int_{0}^{\infty} U_{\ast}(k) k^{4} dk
\int_{-1}^{1} (1-x^{2}) dx \biggl \{
\sum_{m=1}^{\infty} \frac{1}{m^{2}}
\Bigl [ F^{2}(kQx+\pi m)+F^{2}(kQx-\pi m)\Bigr ]
\nonumber\\
&& +F(kQx)\sum_{m=1}^{\infty} \frac{(-1)^{m}}{m^{2}}
\Bigl [ F(kQx+2\pi m)+F(kQx-2\pi m) \Bigr ]
\nonumber\\
&& +2\sum_{m=1}^{\infty} \frac{(-1)^{m}}{m^{2}}
F(kQx+\pi m)F(kQx-\pi m)
-2F^{2}(kQx)\sum_{m=1}^{\infty} \frac{1}{m^{2}}
\biggr \}.
\end{eqnarray*}
Bearing in mind that
\begin{equation}
\sum_{m=1}^{\infty} \frac{1}{m^{2}}=\frac{\pi^{2}}{6},
\end{equation}
we arrive at Eqs. (44) and (45).

We begin with calculation of $s_{1}(z)$.
It follows from Eq. (37) that
\[
F(z+2\pi m)+F(z-2\pi m)
=\frac{1}{z+2\pi m} \sin\Bigl (\frac{z}{2}+\pi m\Bigr )
+\frac{1}{z-2\pi m} \sin\Bigl (\frac{z}{2}-\pi m\Bigr ).
\]
Taking into account that for any integer $m$,
\[
\sin\Bigl (\frac{z}{2}\pm \pi m\Bigr )
=(-1)^{m}\sin\frac{z}{2},
\]
we obtain
\[
F(z+2\pi m)+F(z-2\pi m)
=(-1)^{m}\sin \frac{z}{2} \Bigl (
\frac{1}{z+2\pi m}+\frac{1}{z-2\pi m}\Bigr )
= 2\frac{(-1)^{m}z}{z^{2}-(2\pi m)^{2}}\sin\frac{z}{2}.
\]
This equality together with Eq. (45) implies that
\[
s_{1}(z) =2z\sin\frac{z}{2}\sum_{m=1}^{\infty}
\frac{1}{m^{2}(z^{2}-(2\pi m)^{2})}.
\]
It follows from the identity
\[
\frac{1}{m^{2}(z^{2}-(2\pi m)^{2})}
=\frac{1}{z^{2}}\Bigl [ \frac{1}{m^{2}}
+\frac{(2\pi)^{2}}{z^{2}-(2\pi m)^{2}}\Bigr ]
\]
that
\[
s_{1}(z) =\frac{2}{z}\sin\frac{z}{2}
\biggl [ \sum_{m=1}^{\infty}\frac{1}{m^{2}}
+(2\pi)^{2}\sum_{m=1}^{\infty} \frac{1}{z^{2}-(2\pi m)^{2}}
\biggr ].
\]
Using Eqs. (37) and (D-1), we transform this equality
as follows:
\begin{equation}
s_{1}(z) =2 \pi^{2} F(z) \biggl [ \frac{1}{6}
+4\sum_{m=1}^{\infty} \frac{1}{z^{2}-(2\pi m)^{2}}
\biggr ].
\end{equation}
Setting $z=2z_{1}$, we obtain
\[
4\sum_{m=1}^{\infty} \frac{1}{z^{2}-(2\pi m)^{2}}
=\sum_{m=1}^{\infty} \frac{1}{z_{1}^{2}-(\pi m)^{2}}.
\]
The latter sum is well-known \cite{Man65},
\begin{equation}
\sum_{m=1}^{\infty} \frac{1}{z^{2}-(\pi m)^{2}}
=\frac{1}{2z}\Bigl (\cot z-\frac{1}{z}\Bigr ).
\end{equation}
Substituting expression (D-3) into Eq. (D-2) and
returning to the initial notation, we arrive at
the formula
\begin{equation}
s_{1}(z)=2\pi^{2}F(z)\Bigl [
\frac{1}{6}+\frac{1}{z}\Bigl (\cot
\frac{z}{2}-\frac{2}{z}\Bigr )\Bigr ].
\end{equation}
It follows from Eqs. (37) and (45) that the function
$s_{2}(z)$ reads
\begin{equation}
s_{2}(z) =
\sum_{m=1}^{\infty} \frac{1}{m^{2}}
\Bigl [\frac{\sin^{2}(\frac{z}{2}+\frac{\pi m}{2})}
{(z+\pi m)^{2}}
+\frac{\sin^{2}(\frac{z}{2}-\frac{\pi m}{2})}
{(z-\pi m)^{2}}\Bigr ]
= s_{2}^{(1)}(z)-s_{2}^{(2)}(z)\cos z,
\end{equation}
where
\begin{equation}
s_{2}^{(1)}(z)=\sum_{m=1}^{\infty}
\frac{z^{2}+(\pi m)^{2}}{m^{2}(z^{2}-(\pi m)^{2})^{2}},
\qquad
s_{2}^{(2)}(z)=\sum_{m=1}^{\infty} \frac{(-1)^{m}
(z^{2}+(\pi m)^{2})}{m^{2}(z^{2}-(\pi m)^{2})^{2}}.
\end{equation}
To determine the function $s_{2}^{(1)}(z)$, we
present the first equality in Eq. (D-6) in the form
\begin{equation}
s_{2}^{(1)}(z)=z^{2} \sum_{m=1}^{\infty}
\frac{1}{m^{2}(z^{2}-(\pi m)^{2})^{2}}
+\pi^{2}\sum_{m=1}^{\infty}
\frac{1}{(z^{2}-(\pi m)^{2})^{2}}.
\end{equation}
The first sum in Eq. (D-7) is transformed with
the help of the identity
\begin{equation}
\frac{1}{m^{2}(z^{2}-(\pi m)^{2})^{2}}
=\frac{1}{z^{4}} \Bigl [
\frac{1}{m^{2}}+\frac{\pi^{2}}{z^{2}-(\pi m)^{2}}
+\frac{\pi^{2}z^{2}}{(z^{2}-(\pi m)^{2})^{2}}\Bigr ].
\end{equation}
Combining Eqs. (D-7) and (D-8), we obtain
\begin{equation}
s_{2}^{(1)}(z)=\frac{1}{z^{2}}
\sum_{m=1}^{\infty} \frac{1}{m^{2}}
+\frac{\pi^{2}}{z^{2}}
\sum_{m=1}^{\infty} \frac{1}{z^{2}-(\pi m)^{2}}
+2 \pi^{2} \sum_{m=1}^{\infty}
\frac{1}{(z^{2}-(\pi m)^{2})^{2}}.
\end{equation}
Differentiation of Eq. (D-3) with respect to $z$
implies that
\begin{equation}
\sum_{m=1}^{\infty}
\frac{1}{(z^{2}-(\pi m)^{2})^{2}}
=\frac{1}{4z^{2}}\Bigl (\frac{\cot z}{z}
+\frac{1}{\sin^{2}z}-\frac{2}{z^{2}}\Bigr ).
\end{equation}
Substitution of expressions (D-1), (D-3) and (D-10)
into Eq. (D-9) implies that
\begin{equation}
s_{2}^{(1)}(z)=\frac{\pi^{2}}{2z^{2}}\Bigl (\frac{1}{3}
+2\frac{\cot z}{z}+\frac{1}{\sin^{2}z}-\frac{3}{z^{2}}
\Bigr ).
\end{equation}
According to Eq. (D-6), the function $s_{2}^{(2)}(z)$
is given by
\[
s_{2}^{(2)}(z)=z^{2} \sum_{m=1}^{\infty} \frac{(-1)^{m}}
{m^{2}(z^{2}-(\pi m)^{2})^{2}}
+\pi^{2}\sum_{m=1}^{\infty} \frac{(-1)^{m}}
{(z^{2}-(\pi m)^{2})^{2}}.
\]
Combining this equality with Eq. (D-8), we find that
\begin{equation}
s_{2}^{(2)}(z)=\frac{1}{z^{2}}
\sum_{m=1}^{\infty} \frac{(-1)^{m}}{m^{2}}
+\frac{\pi^{2}}{z^{2}}
\sum_{m=1}^{\infty} \frac{(-1)^{m}}{z^{2}-(\pi m)^{2}}
+2 \pi^{2} \sum_{m=1}^{\infty}
\frac{(-1)^{m}}{(z^{2}-(\pi m)^{2})^{2}}.
\end{equation}
To evaluate the first sum in Eq. (D-12), we present
Eq. (D-1) in the form
\[
\frac{\pi^{2}}{6}
=\sum_{m=1}^{\infty} \frac{1}{(2m-1)^{2}}
+\sum_{m=1}^{\infty} \frac{1}{(2m)^{2}}
=\sum_{m=1}^{\infty} \frac{1}{(2m-1)^{2}}
+\frac{1}{4} \sum_{m=1}^{\infty} \frac{1}{m^{2}}.
\]
Combination of this equality with Eq. (D-1) results in
\begin{equation}
\sum_{m=1}^{\infty} \frac{1}{(2m-1)^{2}}
=\frac{\pi^{2}}{8},
\qquad
\sum_{m=1}^{\infty} \frac{1}{(2m)^{2}}
=\frac{\pi^{2}}{24}.
\end{equation}
Bearing in mind that
\[
\sum_{m=1}^{\infty} \frac{(-1)^{m}}{m^{2}}=
-\sum_{m=1}^{\infty} \frac{1}{(2m-1)^{2}}
+\sum_{m=1}^{\infty} \frac{1}{(2m)^{2}},
\]
we find from Eq. (D-13) that
\begin{equation}
\sum_{m=1}^{\infty} \frac{(-1)^{m}}{m^{2}}
=-\frac{\pi^{2}}{12}.
\end{equation}
The other sum in Eq. (D-12) is well-known \cite{Man65},
\begin{equation}
\sum_{m=1}^{\infty} \frac{(-1)^{m}}{z^{2}-(\pi m)^{2}}
=\frac{1}{2z} \Bigl (\frac{1}{\sin z}-\frac{1}{z}\Bigr ).
\end{equation}
Differentiation of Eq. (D-15) with respect to $z$
results in
\begin{equation}
\sum_{m=1}^{\infty}
\frac{(-1)^{m}}{(z^{2}-(\pi m)^{2})^{2}}
=\frac{1}{4z^{2}} \Bigl (\frac{1}{z\sin z}
+\frac{\cos z}{\sin^{2}z}-\frac{2}{z^{2}}\Bigr ).
\end{equation}
Substitution of expressions (D-14) to (D-16)
into Eq. (D-12) implies that
\[
s_{2}^{(2)}(z)=\frac{\pi^{2}}{2z^{2}}
\Bigl (-\frac{1}{6}+\frac{2}{z\sin z}
+\frac{\cos z}{\sin^{2} z}-\frac{3}{z^{2}}\Bigr ).
\]
Combining this equality with Eqs. (D-5) and (D-11),
we arrive at the formula
\begin{equation}
s_{2}(z)=\frac{\pi^{2}}{2z^{2}}\Bigl (
\frac{8+\cos z}{6}-3\frac{1-\cos z}{z^{2}}\Bigr ).
\end{equation}
It follows from Eqs. (37) and (45) that
\[
s_{3}(z)=\sum_{m=1}^{\infty} \frac{(-1)^{m}}{m^{2}}
\frac{\sin(\frac{z}{2}+\frac{\pi m}{2})\sin(\frac{z}{2}
-\frac{\pi m}{2})}{z^{2}-(\pi m)^{2}}.
\]
Bearing in mind that
\[
\sin(\frac{z}{2}+\frac{\pi m}{2})\sin(\frac{z}{2}
-\frac{\pi m}{2})
=\frac{1}{2}\Bigl (\cos(\pi m)-\cos z\Bigr )
=\frac{1}{2}\Bigl [ (-1)^{m}-\cos z\Bigr ],
\]
we find that
\begin{equation}
s_{3}(z)=\frac{1}{2}\Bigl [ s_{3}^{(1)}(z)
-s_{3}^{(2)}(z)\cos z \Bigr ],
\end{equation}
where
\[
s_{3}^{(1)}(z)=\sum_{m=1}^{\infty}
\frac{1}{m^{2}(z^{2}-(\pi m)^{2})},
\qquad
s_{3}^{(2)}(z)=\sum_{m=1}^{\infty}
\frac{(-1)^{m}}{m^{2}(z^{2}-(\pi m)^{2})}.
\]
Taking into account that
\[
\frac{1}{m^{2}(z^{2}-(\pi m)^{2})}
=\frac{1}{z^{2}}\Bigl [ \frac{1}{m^{2}}
+\frac{\pi^{2}}{z^{2}-(\pi m)^{2}}\Bigr ]
\]
and employing Eqs. (D-1) and (D-3), we obtain
\begin{equation}
s_{3}^{(1)}(z)=\frac{\pi^{2}}{2z^{2}} \Bigl [
\frac{1}{3}+\frac{1}{z}\Bigl (\cot z-\frac{1}{z}\Bigr )
\Bigr ].
\end{equation}
Similarly, we find that
\begin{equation}
s_{3}^{(2)}(z)=\frac{1}{z^{2}} \Bigl [
\sum_{m=1}^{\infty} \frac{(-1)^{m}}{m^{2}}
+\pi^{2} \sum_{m=1}^{\infty}
\frac{(-1)^{m}}{z^{2}-(\pi m)^{2}}\Bigr ].
\end{equation}
Substitution of expressions (D-14) and (D-15) into
Eq. (D-20) implies that
\[
s_{3}^{(2)}(z)=\frac{\pi^{2}}{2z^{2}} \Bigl [
-\frac{1}{6}+\frac{1}{z}
\Bigl (\frac{1}{\sin z}-\frac{1}{z}\Bigr )\Bigr ].
\]
Combining this equality with Eqs. (D-18) and (D-19),
we find that
\begin{equation}
s_{3}(z)= \frac{\pi^{2}}{4z^{2}}
\Bigl ( \frac{2+\cos z}{6}
-\frac{1-\cos z}{z^{2}} \Bigl ).
\end{equation}
Substitution of Eqs. (37), (D-4), (D-17) and (D-21)
into Eq. (45) results in Eq. (46).

It follows from Eqs. (9) and (47) that
\[
S=\frac{\pi^{2}}{b^{2}\tilde{Q}^{2}}
\int_{0}^{1}\frac{1-x^{2}}{x^{2}} dx
\int_{0}^{k_{\ast}} \alpha(kb\tilde{Q}x)k^{2} dk .
\]
Introducing the new variable $z=kb\tilde{Q}x$, we obtain
\[
S=\frac{\pi^{2}}{b^{5}\tilde{Q}^{5}}
\int_{0}^{1} \frac{1-x^{2}}{x^{5}}dx
\int_{0}^{k_{\ast}b\tilde{Q}x} \alpha(z) z^{2} dz.
\]
We now set $y=k_{\ast}b\tilde{Q}x$ and change the order
of integration,
\begin{eqnarray}
S &=& \frac{\pi^{2}k_{\ast}^{4}}{b\tilde{Q}}
\int_{0}^{k_{\ast}b\tilde{Q}}
\Bigl (1-\Bigl (\frac{y}{k_{\ast}b\tilde{Q}}\Bigr )^{2}\Bigr )
\frac{dy}{y^{5}}\int_{0}^{y} \alpha(z) z^{2} dz
\nonumber\\
&=& \frac{\pi^{2}k_{\ast}^{4}}{b\tilde{Q}}
\int_{0}^{k_{\ast}b\tilde{Q}}
\alpha(z) z^{2} dz
\int_{z}^{k_{\ast}b \tilde{Q}} \Bigl (1-\Bigl (
\frac{y}{k_{\ast}b\tilde{Q}}\Bigr )^{2}\Bigr )
\frac{dy}{y^{5}}.
\end{eqnarray}
Substituting the expression
\[
\int_{z}^{k_{\ast}b\tilde{Q}}\Bigl (1
-\Bigl (\frac{y}{k_{\ast}b \tilde{Q}}\Bigr )^{2}\Bigr )
\frac{dy}{y^{5}}
=\frac{1}{4} \Bigl (\frac{1}{z^{2}}
-\frac{1}{(k_{\ast}b\tilde{Q})^{2}} \Bigr )^{2}
\]
into Eq. (D-22), we arrive at Eq. (48).

To transform expression (49) for the function $A_{0}(x)$,
we, first, find from Eq. (46) that
\begin{equation}
\alpha(z)=1+2z^{2}\frac{d}{dz}\Bigl (
\frac{1-\cos z}{z^{3}}\Bigr ).
\end{equation}
Substitution of Eq. (D-23) into Eq. (49) and integration
by parts imply that
\[
A_{0}(x)= \frac{1}{x^{5}}\biggl [
8 \int_{0}^{x}(x^{2}-z^{2}) \frac{1-\cos z}{z^{2}}dz
-4 \int_{0}^{x}(x^{2}-z^{2}) dz \biggr].
\]
Using Eq. (18) and calculating the second integral
in this equality, we obtain
\begin{equation}
A_{0}(x)=\frac{8}{x^{2}}\Bigl [ A(x)-\frac{1}{3}\Bigr ].
\end{equation}
Using Eq. (D-24) and expanding the function $A(x)$ into
the Taylor series in the vicinity of the point $x=0$,
we find that
\begin{equation}
\lim_{x\to 0} A_{0}(x) = 4\frac{d^{2}A}{dx^{2}}(0).
\end{equation}
Differentiation of Eq. (A-8) with respect to $x$
implies that
\[
\frac{d^{2}A}{dx^{2}}(x) = \frac{4}{x^{3}}
\int_{0}^{x} \frac{z-\sin z}{z^{3}} dz
+\frac{-8x +10\sin x-2x\cos x}{x^{5}}.
\]
Applying L'Hospital's rule, we obtain
\begin{equation}
\frac{d^{2}A}{dx^{2}}(0)=-\frac{1}{90}.
\end{equation}
The first equality in Eq. (50) follows from Eqs.
(D-25) and (D-26).
The other equality is a consequence
of Eqs. (19) and (D-24).

\section*{Appendix E}
\renewcommand{\theequation}{E-\arabic{equation}}
\setcounter{equation}{0}

Substitution of expression (D-24) into Eq. (53) implies
that
\begin{equation}
C_{2}=8\delta^{2} \biggl [
\int_{0}^{\infty} A(\frac{x}{\delta})
\exp \Bigl (-\frac{3x^{2}}{2}\Bigr ) dx
-\frac{1}{6}\sqrt{\frac{2\pi}{3}}\biggr ].
\end{equation}
Our aim now is to prove that
the first term in the square brackets
\begin{equation}
I= \int_{0}^{\infty} A(\frac{x}{\delta})
\exp \Bigl (-\frac{3x^{2}}{2}\Bigr ) dx
\end{equation}
is small compared with unity.
It follows from Eq. (A-8) that for any $y$,
\begin{equation}
A(y)-\frac{2}{y}\int_{0}^{\infty} a(z) dz
=-\frac{2}{y}\int_{y}^{\infty} a(z)dz.
\end{equation}
Beaing in mind that
\[
0\leq a(z)=\frac{1}{z^{2}}\Bigl (1-\frac{\sin z}{z}\Bigr )
\leq \frac{2}{z^{2}},
\]
we find from Eq. (E-3) that
\begin{equation}
\Bigl |A(y)-\frac{2}{y}\int_{0}^{\infty} a(z) dz\Bigr |
\leq \frac{4}{y}\int_{y}^{\infty} \frac{dz}{z^{2}}
=\frac{4}{y^{2}}.
\end{equation}
Setting $y=x/\delta$ in Eq. (E-2), we obtain
\begin{equation}
I=\delta\int_{0}^{\infty} A(y)\exp
\Bigl (-\frac{3\delta^{2}y^{2}}{2}\Bigr ) dy
=\delta (I_{1}+I_{2}+I_{3}),
\end{equation}
where
\begin{eqnarray}
I_{1} &=& \int_{0}^{1} A(y) \exp \Bigl (
-\frac{3\delta^{2}y^{2}}{2}\Bigr ) dy,
\nonumber\\
I_{2} &=& \int_{1}^{\infty} \Bigl (A(y)
-\frac{2}{y} \int_{0}^{\infty} a(z)dz \Bigr )
\exp \Bigl (-\frac{3\delta^{2}y^{2}}{2}\Bigr ) dy ,
\nonumber\\
I_{3} &=& 2 \int_{0}^{\infty} a(z)dz
\int_{1}^{\infty} \exp \Bigl (-\frac{3\delta^{2}y^{2}}{2}
\Bigr ) \frac{dy}{y}.
\end{eqnarray}
It follows from Eqs. (19) and (E-6) that
\begin{equation}
|I_{1}| \leq \int_{0}^{1} A(y) dy \leq \frac{1}{3}.
\end{equation}
Equations (E-4) and (E-6) imply that
\begin{equation}
|I_{2}| \leq 4 \int_{1}^{\infty}
\exp \Bigl (-\frac{3\delta^{2}y^{2}}{2}\Bigr )
\frac{dy}{y^{2}}
\leq 4 \int_{1}^{\infty} \frac{dy}{y^{2}}=4.
\end{equation}
According to Eqs. (B-14) and (E-6),
\[
I_{3}=\frac{\pi}{2}\int_{1}^{\infty}
\exp \Bigl (-\frac{3\delta^{2}y^{2}}{2}
\Bigr ) \frac{dy}{y}
=\frac{\pi}{2}\int_{\delta}^{\infty}
\exp \Bigl (-\frac{3x^{2}}{2}\Bigr ) \frac{dx}{x},
\]
where we return to the variable $x$.
For $\delta < 1$, this equality reads
\begin{equation}
I_{3}=\frac{\pi}{2}(I_{3}^{(1)}+I_{3}^{(2)}),
\end{equation}
where
\[
I_{3}^{(1)}=\int_{\delta}^{1}
\exp \Bigl (-\frac{3x^{2}}{2}\Bigr ) \frac{dx}{x},
\qquad
I_{3}^{(2)}=\int_{1}^{\infty}\exp
\Bigl (-\frac{3x^{2}}{2}\Bigr ) \frac{dx}{x}.
\]
It follows from these equalities that
\begin{equation}
|I_{3}^{(1)}|\leq \int_{\delta}^{1}\frac{dx}{x}=-\ln \delta,
\qquad
|I_{3}^{(2)}|\leq \int_{1}^{\infty} \exp
\Bigl (-\frac{3x^{2}}{2}\Bigr ) dx
\leq \int_{0}^{\infty} \exp
\Bigl (-\frac{3x^{2}}{2}\Bigr ) dx
=\sqrt{\frac{\pi}{6}}.
\end{equation}
Combining Eq. (E-5) with Eqs. (E-7) to (E-10),
we arrive at the formula
\begin{equation}
|I|\leq \delta \Bigl [ \frac{13}{3}
+\frac{\pi}{2}\Bigl (\sqrt{\frac{\pi}{6}}
-\ln \delta\Bigr )\Bigr ],
\end{equation}
which completes the proof.

It follows from Eqs. (57) and (D-24) that
\begin{equation}
B_{2} = 8\delta^{2}\Bigl [ \int_{0}^{\infty} A(\frac{x}{\delta})
\exp \Bigl (-\frac{3x^{2}}{2}\Bigr ) x^{2} dx
-\frac{1}{3} \int_{0}^{\infty}
\exp \Bigl (-\frac{3x^{2}}{2}\Bigr ) x^{2} dx
\Bigr ] .
\end{equation}
According to Eq. (25), the first integral in Eq. (E-12)
coincides with $C_{1}$.
The other integral is given by Eq. (B-1).
Substitution of these expressions into Eq. (E-12)
results in
\[
B_{2}=8\delta^{2} \Bigl (C_{1}-\frac{1}{18}
\sqrt{\frac{2\pi}{3}}\Bigr ).
\]
Combining this equality with Eq. (26) and disregarding
terms beyond the second order of smallness,
we arrive at Eq. (58).

\section*{Appendix F}
\renewcommand{\theequation}{F-\arabic{equation}}
\setcounter{equation}{0}

To find $\Lambda_{1}$, we formally expand
the denominator in the first equality in Eq. (64)
into the Taylor series in $\mu_{1}=-\mu$,
\begin{equation}
\frac{A_{1}(y)}{1-\mu_{1}A_{1}(y)}=A_{1}(y)\Bigl [
1+\mu_{1}A_{1}(y)+\sum_{m=2}^{\infty}
\Bigl (\mu_{1}A_{1}(y)\Bigr )^{m} \Bigr ],
\end{equation}
and evaluate the integrals of appropriate terms
in this sum.
The integral
\[
\Lambda_{1}^{(1)}=\int_{0}^{\infty} A_{1}(\frac{x}{\delta})
\exp \Bigl (-\frac{3x^{2}}{2}\Bigr ) x^{2}dx
\]
was calculated previously.
It follows from Eqs. (53), (54) and (62) that the
leading term in the expression for $\Lambda_{1}^{(1)}$
is given by
\begin{equation}
\Lambda_{1}^{(1)}=\frac{2}{27}\sqrt{\frac{2\pi}{3}}
\delta^{2}.
\end{equation}
To find the coefficient
\[
\Lambda_{1}^{(2)}=\int_{0}^{\infty}
A_{1}^{2}(\frac{x}{\delta})
\exp \Bigl (-\frac{3x^{2}}{2}\Bigr ) x^{2}dx,
\]
we substitute expressions (62) and (D-24) into
this equality and obtain
\begin{equation}
\Lambda_{1}^{(2)}=\frac{2}{81}\delta^{2}
\int_{0}^{\infty} A_{0}(\frac{x}{\delta})
\Bigl [A(\frac{x}{\delta})-\frac{1}{3}\Bigr ]
\exp \Bigl (-\frac{3x^{2}}{2}\Bigr ) dx.
\end{equation}
The function
\[
\alpha_{1}(x)=A_{0}(x)\Bigl [A(x)-\frac{1}{3}\Bigr ]
\]
is even and continuous, it has the zero limit at $x\to 0$
and vanishes being proportional to $x^{-2}$
at $x\to \infty$.
This means that the Fourier transform $\hat{\alpha}_{1}(s)$
of this function exists.
Presenting Eq. (F-3) in the form
\[
\Lambda_{1}^{(2)}=\frac{\delta^{2}}{162\pi}
\int_{-\infty}^{\infty}
\exp \Bigl (-\frac{3x^{2}}{2}\Bigr ) dx
\int_{-\infty}^{\infty} \hat{\alpha}_{1}(s)
\exp \Bigl (-\frac{\imath s x}{\delta}\Bigr )ds
\]
and evaluating the integral with the use of Eq. (B-7),
we arrive at
\[
\Lambda_{1}^{(2)}=\frac{\delta^{3}}{81\sqrt{2\pi}}
\int_{-\infty}^{\infty} \hat{\alpha}_{1}(\delta z \sqrt{3})
\exp \Bigl (-\frac{z^{2}}{2}\Bigr ) dz.
\]
The leading term in this expression is obtained
when the function $\hat{\alpha}_{1}(s)$ is expanded
into the Taylor series in the vicinity of $s=0$
and only the first term in the series is taken into
account,
\begin{equation}
\Lambda_{1}^{(2)}=\frac{\delta^{3}}{81}\hat{\alpha}_{1}(0)
=\frac{2\delta^{3}}{81}\int_{0}^{\infty}
A_{0}(x)\Bigl [ A(x)-\frac{1}{3}\Bigr ] dx.
\end{equation}
The integrals
\[
\Lambda_{1}^{(m)}=\int_{0}^{\infty}
A_{1}^{m}(\frac{x}{\delta})
\exp \Bigl (-\frac{3x^{2}}{2}\Bigr ) x^{2}dx
\]
with $m\geq 2$ are estimated by using a standard
approach.
Setting
\begin{equation}
\alpha_{m}(x)=A_{1}^{m}(x),
\end{equation}
we re-write this equality in the form
\begin{equation}
\Lambda_{1}^{(m)}=\int_{0}^{\infty}
\alpha_{m}(\frac{x}{\delta})
\exp \Bigl (-\frac{3x^{2}}{2} \Bigr ) x^{2} dx.
\end{equation}
It follows from Eqs. (62) and (D-24) that $A_{1}^{m}(x)$
vanishes at $x\to \infty$ being proportional to $x^{-2m}$.
This means that for any $m \geq 2$, the second derivative
with respect to $s$ exists of the Fourier transform
\begin{equation}
\hat{\alpha}_{m}(s)=\int_{-\infty}^{\infty} \alpha_{m}(x)
\exp (\imath sx)dx
\end{equation}
of the function $\alpha_{m}(x)$.
Bearing in mind that the function $\alpha_{m}(x)$ is even,
substituting the expression
\[
\alpha_{m}(\frac{x}{\delta})=\frac{1}{2\pi}
\int_{-\infty}^{\infty} \hat{\alpha}_{m}(s)
\exp (-\frac{\imath sx}{\delta})ds
\]
into Eq. (F-6) and changing the order of
integration, we find that
\[
\Lambda_{1}^{(m)}=\frac{1}{4\pi} \int_{-\infty}^{\infty}
\hat{\alpha}_{m}(s)ds
\int_{-\infty}^{\infty}
\exp \Bigl [-\Bigl (\frac{3x^{2}}{2}
+\frac{\imath sx}{\delta} \Bigr ) \Bigr ]
x^{2} dx.
\]
Using Eq. (B-19) for the internal integral
and setting
$z=s/(\delta\sqrt{3})$, we arrive at
\begin{equation}
\Lambda_{1}^{(m)}=\frac{\delta}{6\sqrt{2\pi}}
\int_{-\infty}^{\infty} \hat{\alpha}_{m}(\delta z \sqrt{3})
\exp \Bigl (-\frac{z^{2}}{2}\Bigr )
(1-z^{2})dz.
\end{equation}
Expanding the even function $\hat{\alpha}_{m}(s)$ into the
Taylor series in the vicinity of the point $s=0$
and neglecting terms beyond the second order of smallness,
we find that
\[
\hat{\alpha}_{m}(\delta z \sqrt{3})
=\hat{\alpha}_{m}(0)+\frac{3\delta^{2}}{2}
\frac{d^{2}\hat{\alpha}_{m}}{ds^{2}}(0)z^{2}.
\]
We combine this equality with Eq. (F-8),
use Eq. (B-21), and obtain
\begin{equation}
\Lambda_{1}^{(m)}=\frac{\delta^{3}}{4\sqrt{2\pi}}
\frac{d^{2}\hat{\alpha}_{m}}{ds^{2}}(0)
\int_{-\infty}^{\infty}
\exp \Bigl (-\frac{z^{2}}{2}\Bigr )(1-z^{2})z^{2} dz
=-\frac{\delta^{3}}{2}
\frac{d^{2}\hat{\alpha}_{m}}{ds^{2}}(0).
\end{equation}
It follows from Eq. (F-7) that
\[
\frac{d^{2}\hat{\alpha}_{m}}{ds^{2}}(0)
=-2\int_{0}^{\infty} \alpha_{m}(x)x^{2} dx.
\]
This equality together with Eqs. (F-5) and (F-9)
yields
\begin{equation}
\Lambda_{1}^{(m)}=\delta^{3}
\int_{0}^{\infty} A_{1}^{m}(x)x^{2} dx.
\end{equation}
Substitution of expressions (F-1), (F-2), (F-4)
and (F-10) into Eq. (64) results in
\[
\Lambda_{1}=\delta^{2} \biggl [ \frac{2}{27}
\sqrt{\frac{2\pi}{3}}+\frac{2}{81}\mu_{1}\delta
\int_{0}^{\infty} A_{0}(x)
\Bigl ( A(x)-\frac{1}{3}\Bigr )dx
+\frac{\delta}{\mu_{1}} \int_{0}^{\infty}
\sum_{m=3}^{\infty} \mu_{1}^{m}A_{1}^{m}(x)x^{2}dx
\biggr ].
\]
Taking into account that
\begin{equation}
\sum_{m=3}^{\infty} \Bigl (\mu_{1}A_{1}(x)\Bigr )^{m}
=\frac{\mu_{1}^{3}A_{1}^{3}(x)}{1-\mu_{1}A_{1}(x)},
\end{equation}
and returning to the initial notation, we find that
\[
\Lambda_{1}=\delta^{2} \biggl [ \frac{2}{27}
\sqrt{\frac{2\pi}{3}}+\frac{4}{9}\mu \delta
\int_{0}^{\infty} A_{1}(x)
\Bigl ( A(x)-\frac{1}{3}\Bigr )dx
+\mu^{2}\delta\int_{0}^{\infty}
\frac{A_{1}^{3}(x)x^{2}}{1+\mu A_{1}(x)}dx
\biggr ].
\]
Combining the integral terms and utilizing
Eqs. (62) and (D-24), we obtain
\begin{equation}
\Lambda_{1}=\delta^{2} \biggl [ \frac{2}{27}
\sqrt{\frac{2\pi}{3}}-\mu \delta
\int_{0}^{\infty} \frac{A_{1}^{2}(x)x^{2}}{1
+\mu A_{1}(x)}dx \biggr ].
\end{equation}
As all integrals in Eq. (F-12) converge, it can be shown
that the above transformations are correct for an
arbitrary (not necessary small) $\mu$.

Equation (F-12) implies that the leading term
in the expression for $\Lambda_{1}$ is (at least)
of order of $\delta^{2}$.
As the quantity $\Lambda_{2}$ is included into
Eq. (63) with the pre-factor $\delta^{2}$, in order
to disregard this term it suffices to show that
$\Lambda_{2}$ is small compared with unity.
Taking into account that the function $A_{1}(x)$ is
non-negative, we write
\begin{equation}
\frac{A(x)}{1+\mu A_{1}(x)}\leq A(x) .
\end{equation}
This equality together with Eqs. (25) and (64)
implies that
\[
0\leq \Lambda_{2}\leq \int_{0}^{\infty}
A(\frac{x}{\delta})\exp \Bigl (-\frac{3x^{2}}{2}
\Bigr ) x^{2} dx =C_{1}.
\]
It follows from this inequality and Eq. (26)
that the contribution of $\Lambda_{2}$ into
Eq. (63) is negligible compared with $\Lambda_{1}$.
This conclusion together with Eqs. (63) and (F-12)
implies that
\[
C=\Bigl (\frac{3}{2\pi b^{2}}\Bigr )^{\frac{3}{2}}
\biggl \{1-\mu \delta^{2}\biggl [ \frac{4}{9}
-6\sqrt{\frac{3}{2 \pi}}\mu\delta
\int_{0}^{\infty} \frac{A_{1}^{2}(x)x^{2}}{1
+\mu A_{1}(x)}dx \biggr ] \biggr \}^{-1}.
\]
Equation (65) follows from this equality and Eq. (62).

To find the coefficient $\Gamma_{1}$, we evaluate
the integrals of appropriate terms in expansion (F-1).
The integral
\[
\Gamma_{1}^{(1)}=\int_{0}^{\infty} A_{1}(\frac{x}{\delta})
\exp \Bigl (-\frac{3x^{2}}{2}\Bigr ) x^{4}dx
\]
was calculated previously.
It follows from Eqs. (57), (58) and (62) that the
leading term in the expression for $\Gamma_{1}^{(1)}$
reads
\begin{equation}
\Gamma_{1}^{(1)}=\frac{2}{81}\sqrt{\frac{2\pi}{3}}
\delta^{2}.
\end{equation}
Using Eqs. (62) and (D-24), we transform the
coefficient
\[
\Gamma_{1}^{(2)}=\int_{0}^{\infty} A_{1}^{2}(\frac{x}{\delta})
\exp \Bigl (-\frac{3x^{2}}{2}\Bigr ) x^{4}dx
\]
as follows:
\[
\Gamma_{1}^{(2)}
= \frac{16}{81}\delta^{4}\biggl [
\int_{0}^{\infty} A^{2}(\frac{x}{\delta})
\exp \Bigl (-\frac{3x^{2}}{2}\Bigr ) dx
-\frac{2}{3}\int_{0}^{\infty} A(\frac{x}{\delta})
\exp \Bigl (-\frac{3x^{2}}{2}\Bigr ) dx
+\frac{1}{9}\int_{0}^{\infty}
\exp \Bigl (-\frac{3x^{2}}{2}\Bigr ) dx\biggr ].
\]
The first term in the square brackets is of order
of $\delta$, because the Fourier transform
of the function $A^{2}(x)$ exists.
The second term is estimated in Eq. (E-11), where
it is shown that it is of order of $\delta|\ln\delta|$.
The last term is calculated explicitly, and it is of order
of unity.
Neglecting small contributions into the coefficient
$\Gamma_{1}^{(2)}$, we obtain
\begin{equation}
\Gamma_{1}^{(2)}=\frac{16}{729}\sqrt{\frac{\pi}{6}}
\delta^{4}.
\end{equation}
For any integer $m\geq 3$, the coefficient
\[
\Gamma_{1}^{(m)}=\int_{0}^{\infty} A_{1}^{m}
(\frac{x}{\delta}) \exp \Bigl (-\frac{3x^{2}}{2}\Bigr )
x^{4}dx
\]
reads
\[
\Gamma_{1}^{(m)}=\delta^{4} \int_{0}^{\infty}
\beta_{m}(\frac{x}{\delta})
\exp \Bigl (-\frac{3x^{2}}{2}\Bigr )dx,
\]
where
$\beta_{m}(x)=A_{1}^{m}(x)x^{4}$.
As $A_{1}^{m}(x)$ is an even continuous function,
and it decreases being proportional to $x^{-2m}$
as $x\to \infty$, the Fourier transform
$\hat{\beta}_{m}(s)$ exists of the function
$\beta_{m}(x)$.
Evaluating the integral by analogy with Eq. (B-3),
we find the leading term in the expression for
$\Gamma_{1}^{(m)}$,
\begin{equation}
\Gamma_{1}^{(m)}=\delta^{5}\int_{0}^{\infty}
A_{1}^{(m)}(x) x^{4} dx.
\end{equation}
It follows from Eqs. (F-1) and (F-14) to (F-16) that
\[
\Gamma_{1} = \frac{2}{81} \sqrt{\frac{2\pi}{3}}
\delta^{2}\biggl [ 1+\frac{4}{9} \mu_{1}\delta^{2}
+\frac{81}{2}\sqrt{\frac{3}{2\pi}}\frac{\delta^{3}}{\mu_{1}}
\int_{0}^{\infty} \sum_{m=3}^{\infty}
\Bigl (\mu_{1} A_{1}(x)\Bigr )^{m} x^{4}dx\biggr ].
\]
Combining this equality with Eq. (F-11) and returning
to the initial notation, we arrive at the formula
\begin{equation}
\Gamma_{1}= \frac{2}{81} \sqrt{\frac{2\pi}{3}}
\delta^{2}\biggl [ 1-\frac{4}{9} \mu \delta^{2}
+\frac{81}{2}\sqrt{\frac{3}{2\pi}}\mu^{2} \delta^{3}
\int_{0}^{\infty} \frac{A_{1}^{3}(x) x^{4}}{1
+\mu A_{1}(x)}dx \biggr ] .
\end{equation}
To assess the coefficient $\Gamma_{2}$, we use Eqs.
(25), (26) and (F-13) and obtain (up to terms of
higher order of smallness)
\begin{equation}
0\leq \Gamma_{2}\leq
\int_{0}^{\infty} A(\frac{x}{\delta})
\Bigl (-\frac{3x^{2}}{2}\Bigr )x^{4}dx
=\frac{\pi \delta}{9},
\end{equation}
which implies that the contribution of $\Gamma_{2}$
into the mean square end-to-end distance is
negligible compared with that of $\Gamma_{1}$
[the integral $\Gamma_{2}$ is multiplied
by $\delta^{2}$ in Eq. (66)].
Substitution of Eq. (F-17) into Eq. (66) results in
\[
\Bigl (\frac{B}{b}\Bigr )^{2}=\Bigl (
\frac{2\pi b^{2}}{3}\Bigr )^{\frac{3}{2}}
C \biggl \{1-\frac{4}{27}\mu\delta^{2}
\biggl [1-\frac{4}{9}\mu \delta^{2}
+\frac{81}{2}\sqrt{\frac{3}{2\pi}}\mu^{2}\delta^{3}
\int_{0}^{\infty} \frac{A_{1}^{3}(x)x^{4}}{1+\mu A_{1}(x)}dx
\biggr ]\biggr \}.
\]
Combining this equality with Eq. (62), we conclude that
the second term in the square brackets is negligible
compared with the first.
Neglecting this contribution and returning to the
initial notation, we arrive at Eq. (68).

To assess the integral terms in Eqs. (65) and (69),
we introduce the function
\begin{equation}
D(\mu)=\mu \int_{0}^{\infty}
\frac{A_{1}^{2}(x)x^{2}}{1+\mu A_{1}(x)}dx.
\end{equation}
If the function $A_{1}(x)x^{2}$ had been integrable,
we could use the inequality
\[
\frac{\mu A_{1}(x)}{1+\mu A_{1}(x)}\leq 1
\]
to find that
$D(\mu)\leq \int_{0}^{\infty} A_{1}(x)x^{2}dx
<\infty$
for any positive $\mu$.
As this is not the case, more sophisticated estimates
are needed.
It follows from Eqs. (62) and (D-24) that
\[
A_{1}(x)=\frac{4}{9x^{2}}\Bigl (\frac{1}{3}-A(x)\Bigr ).
\]
Substitution of this expression into Eq. (F-19) yields
\begin{equation}
D(\mu)=\Bigl (\frac{4}{9}\Bigr )^{2}\mu\int_{0}^{\infty}
\frac{(\frac{1}{3}-A(x))^{2}}{x^{2}
+\frac{4}{9}\mu (\frac{1}{3}-A(x))}dx.
\end{equation}
As the function $A(x)$ monotonically decreases with $x$
and vanishes at $x\to \infty$, there exists an $x_{0}$
such that
\begin{equation}
\frac{1}{6}\leq A(x)\leq \frac{1}{3}
\quad(0\leq x\leq x_{0}),
\qquad
0< A(x)<\frac{1}{6}
\quad
(x_{0}<x<\infty).
\end{equation}
We present the integral in Eq. (F-20) as the sum of
two integrals: from zero to $x_{0}$, and from $x_{0}$ to
infinity, and evaluate appropriate integrals separately,
\begin{eqnarray}
D(\mu) &=& \Bigl (\frac{4}{9}\Bigr )^{2}
\Bigl [D_{1}(\mu)+D_{2}(\mu)\Bigr ],
\qquad
D_{1}(\mu) = \mu\int_{0}^{x_{0}}
\frac{(\frac{1}{3}-A(x))^{2}}{x^{2}
+\frac{4}{9}\mu (\frac{1}{3}-A(x))}dx,
\nonumber\\
D_{2}(\mu) &=& \mu\int_{x_{0}}^{\infty}
\frac{(\frac{1}{3}-A(x))^{2}}{x^{2}
+\frac{4}{9}\mu (\frac{1}{3}-A(x))}dx .
\end{eqnarray}
Taking into account that in the interval $x\in [0,x_{0}]$,
\[
x^{2}+\frac{4}{9}\mu \Bigl (\frac{1}{3}-A(x)\Bigr )
\geq \frac{4}{9}\mu \Bigl (\frac{1}{3}-A(x)\Bigr ),
\qquad
\frac{1}{3}-A(x)\leq \frac{1}{6},
\]
the function $D_{1}(\mu)$ is estimated as follows:
\begin{eqnarray}
D_{1}(\mu)\leq \frac{9}{4}\int_{0}^{x_{0}}
\Bigl (\frac{1}{3}-A(x)\Bigr ) dx
\leq \frac{3}{8}x_{0}.
\end{eqnarray}
Using the estimates in the interval $x\in (x_{0},\infty)$,
\[
x^{2}+\frac{4}{9}\mu \Bigl (\frac{1}{3}-A(x)\Bigr )
\geq x^{2}+\frac{2}{27}\mu,
\qquad
\frac{1}{3}-A(x)\leq \frac{1}{3},
\]
the function $D_{2}(\mu)$ is evaluated as
\begin{equation}
D_{2}(\mu)\leq \frac{\mu}{9} \int_{x_{0}}^{\infty}
\frac{dx}{x^{2}+\frac{2}{27}\mu}
\leq \frac{\mu}{9} \int_{0}^{\infty}
\frac{dx}{x^{2}+\frac{2}{27}\mu}
=\frac{\mu}{9}\sqrt{\frac{27}{2\mu}}
\int_{0}^{\infty} \frac{dy}{y^{2}+1}
=\frac{\pi}{2}\sqrt{\frac{\mu}{6}}.
\end{equation}
It follows from Eqs. (F-22) to (F-24) that there are
positive constants $D^{(0)}$ and $D^{(1)}$ such that
for any $\mu$,
\[
D(\mu)\leq D^{(0)}+D^{(1)}\sqrt{\mu}
=D^{(0)}+D^{(1)}\frac{\sqrt{\varepsilon}}{\delta\sqrt{2\pi}},
\]
where notation (62) is employed.
This formula implies that the integral
term in Eqs. (65) and (69) obeys the inequality
\begin{equation}
\frac{\varepsilon}{\delta}\int_{0}^{\infty}
\frac{A_{1}^{2}(x)x^{2}}{1+\mu A_{1}(x)}dx
=2\pi \delta \mu \int_{0}^{\infty}
\frac{A_{1}^{2}(x)x^{2}}{1+\mu A_{1}(x)}dx
\leq 2\pi D^{(0)}\delta +D^{(1)}\sqrt{2\pi \varepsilon},
\end{equation}
and it can be disregarded compared with unity.

\newpage
\setlength{\unitlength}{0.75 mm}
\begin{figure}[tbh]
\begin{center}
\begin{picture}(100,100)
\put(0,0){\framebox(100,100)}
\multiput(10,0)(10,0){9}{\line(0,1){2}}
\multiput(0,10)(0,10){9}{\line(1,0){2}}
\multiput(100,20)(0,20){4}{\line(-1,0){2}}
\put(0,-9){0.0}
\put(89,-9){100.0}
\put(50,-9){$x$}
\put(-10,0){0.0}
\put(-10,96){0.2}
\put(-10,70){$A$}
\put(103,0){$-0.05$}
\put(103,96){0.0}
\put(103,70){$A_{0}$}
\put(80, 8){$A$}
\put(80,92){$A_{0}$}

\put(   0.00,   83.33){\circle*{0.8}}
\put(   0.40,   83.11){\circle*{0.8}}
\put(   0.67,   82.71){\circle*{0.8}}
\put(   0.86,   82.31){\circle*{0.8}}
\put(   1.02,   81.91){\circle*{0.8}}
\put(   1.16,   81.51){\circle*{0.8}}
\put(   1.28,   81.10){\circle*{0.8}}
\put(   1.39,   80.70){\circle*{0.8}}
\put(   1.50,   80.30){\circle*{0.8}}
\put(   1.60,   79.90){\circle*{0.8}}
\put(   1.70,   79.49){\circle*{0.8}}
\put(   1.79,   79.09){\circle*{0.8}}
\put(   1.87,   78.69){\circle*{0.8}}
\put(   1.96,   78.29){\circle*{0.8}}
\put(   2.04,   77.88){\circle*{0.8}}
\put(   2.12,   77.48){\circle*{0.8}}
\put(   2.19,   77.08){\circle*{0.8}}
\put(   2.27,   76.67){\circle*{0.8}}
\put(   2.34,   76.27){\circle*{0.8}}
\put(   2.41,   75.87){\circle*{0.8}}
\put(   2.48,   75.46){\circle*{0.8}}
\put(   2.55,   75.06){\circle*{0.8}}
\put(   2.62,   74.66){\circle*{0.8}}
\put(   2.69,   74.25){\circle*{0.8}}
\put(   2.75,   73.85){\circle*{0.8}}
\put(   2.82,   73.44){\circle*{0.8}}
\put(   2.88,   73.04){\circle*{0.8}}
\put(   2.95,   72.64){\circle*{0.8}}
\put(   3.01,   72.24){\circle*{0.8}}
\put(   3.07,   71.83){\circle*{0.8}}
\put(   3.13,   71.43){\circle*{0.8}}
\put(   3.19,   71.03){\circle*{0.8}}
\put(   3.25,   70.63){\circle*{0.8}}
\put(   3.31,   70.23){\circle*{0.8}}
\put(   3.37,   69.82){\circle*{0.8}}
\put(   3.43,   69.42){\circle*{0.8}}
\put(   3.49,   69.01){\circle*{0.8}}
\put(   3.55,   68.61){\circle*{0.8}}
\put(   3.61,   68.21){\circle*{0.8}}
\put(   3.67,   67.80){\circle*{0.8}}
\put(   3.72,   67.40){\circle*{0.8}}
\put(   3.78,   66.99){\circle*{0.8}}
\put(   3.84,   66.59){\circle*{0.8}}
\put(   3.90,   66.19){\circle*{0.8}}
\put(   3.95,   65.79){\circle*{0.8}}
\put(   4.01,   65.38){\circle*{0.8}}
\put(   4.07,   64.98){\circle*{0.8}}
\put(   4.12,   64.58){\circle*{0.8}}
\put(   4.18,   64.17){\circle*{0.8}}
\put(   4.24,   63.77){\circle*{0.8}}
\put(   4.30,   63.36){\circle*{0.8}}
\put(   4.35,   62.96){\circle*{0.8}}
\put(   4.41,   62.55){\circle*{0.8}}
\put(   4.47,   62.15){\circle*{0.8}}
\put(   4.52,   61.75){\circle*{0.8}}
\put(   4.58,   61.34){\circle*{0.8}}
\put(   4.64,   60.94){\circle*{0.8}}
\put(   4.69,   60.54){\circle*{0.8}}
\put(   4.75,   60.14){\circle*{0.8}}
\put(   4.81,   59.73){\circle*{0.8}}
\put(   4.87,   59.33){\circle*{0.8}}
\put(   4.93,   58.92){\circle*{0.8}}
\put(   4.98,   58.52){\circle*{0.8}}
\put(   5.04,   58.12){\circle*{0.8}}
\put(   5.10,   57.71){\circle*{0.8}}
\put(   5.16,   57.31){\circle*{0.8}}
\put(   5.22,   56.91){\circle*{0.8}}
\put(   5.28,   56.51){\circle*{0.8}}
\put(   5.34,   56.10){\circle*{0.8}}
\put(   5.40,   55.70){\circle*{0.8}}
\put(   5.46,   55.29){\circle*{0.8}}
\put(   5.52,   54.89){\circle*{0.8}}
\put(   5.58,   54.49){\circle*{0.8}}
\put(   5.64,   54.08){\circle*{0.8}}
\put(   5.71,   53.68){\circle*{0.8}}
\put(   5.77,   53.28){\circle*{0.8}}
\put(   5.83,   52.87){\circle*{0.8}}
\put(   5.90,   52.47){\circle*{0.8}}
\put(   5.96,   52.07){\circle*{0.8}}
\put(   6.02,   51.66){\circle*{0.8}}
\put(   6.09,   51.26){\circle*{0.8}}
\put(   6.16,   50.86){\circle*{0.8}}
\put(   6.22,   50.45){\circle*{0.8}}
\put(   6.29,   50.05){\circle*{0.8}}
\put(   6.36,   49.65){\circle*{0.8}}
\put(   6.43,   49.25){\circle*{0.8}}
\put(   6.50,   48.84){\circle*{0.8}}
\put(   6.57,   48.44){\circle*{0.8}}
\put(   6.64,   48.04){\circle*{0.8}}
\put(   6.71,   47.64){\circle*{0.8}}
\put(   6.78,   47.24){\circle*{0.8}}
\put(   6.86,   46.83){\circle*{0.8}}
\put(   6.93,   46.43){\circle*{0.8}}
\put(   7.01,   46.03){\circle*{0.8}}
\put(   7.08,   45.63){\circle*{0.8}}
\put(   7.16,   45.23){\circle*{0.8}}
\put(   7.24,   44.82){\circle*{0.8}}
\put(   7.32,   44.42){\circle*{0.8}}
\put(   7.41,   44.02){\circle*{0.8}}
\put(   7.49,   43.62){\circle*{0.8}}
\put(   7.57,   43.21){\circle*{0.8}}
\put(   7.66,   42.81){\circle*{0.8}}
\put(   7.75,   42.41){\circle*{0.8}}
\put(   7.84,   42.01){\circle*{0.8}}
\put(   7.93,   41.61){\circle*{0.8}}
\put(   8.02,   41.20){\circle*{0.8}}
\put(   8.12,   40.80){\circle*{0.8}}
\put(   8.21,   40.40){\circle*{0.8}}
\put(   8.31,   39.99){\circle*{0.8}}
\put(   8.41,   39.59){\circle*{0.8}}
\put(   8.52,   39.19){\circle*{0.8}}
\put(   8.62,   38.79){\circle*{0.8}}
\put(   8.73,   38.39){\circle*{0.8}}
\put(   8.84,   37.98){\circle*{0.8}}
\put(   8.95,   37.58){\circle*{0.8}}
\put(   9.06,   37.18){\circle*{0.8}}
\put(   9.18,   36.78){\circle*{0.8}}
\put(   9.30,   36.38){\circle*{0.8}}
\put(   9.43,   35.98){\circle*{0.8}}
\put(   9.55,   35.58){\circle*{0.8}}
\put(   9.68,   35.17){\circle*{0.8}}
\put(   9.81,   34.77){\circle*{0.8}}
\put(   9.95,   34.37){\circle*{0.8}}
\put(  10.09,   33.97){\circle*{0.8}}
\put(  10.23,   33.57){\circle*{0.8}}
\put(  10.38,   33.17){\circle*{0.8}}
\put(  10.53,   32.77){\circle*{0.8}}
\put(  10.68,   32.37){\circle*{0.8}}
\put(  10.84,   31.96){\circle*{0.8}}
\put(  11.00,   31.56){\circle*{0.8}}
\put(  11.16,   31.16){\circle*{0.8}}
\put(  11.33,   30.76){\circle*{0.8}}
\put(  11.50,   30.36){\circle*{0.8}}
\put(  11.68,   29.96){\circle*{0.8}}
\put(  11.86,   29.56){\circle*{0.8}}
\put(  12.05,   29.16){\circle*{0.8}}
\put(  12.24,   28.76){\circle*{0.8}}
\put(  12.44,   28.36){\circle*{0.8}}
\put(  12.64,   27.96){\circle*{0.8}}
\put(  12.85,   27.55){\circle*{0.8}}
\put(  13.06,   27.15){\circle*{0.8}}
\put(  13.28,   26.75){\circle*{0.8}}
\put(  13.50,   26.35){\circle*{0.8}}
\put(  13.73,   25.95){\circle*{0.8}}
\put(  13.97,   25.55){\circle*{0.8}}
\put(  14.21,   25.15){\circle*{0.8}}
\put(  14.47,   24.75){\circle*{0.8}}
\put(  14.73,   24.35){\circle*{0.8}}
\put(  15.00,   23.95){\circle*{0.8}}
\put(  15.28,   23.55){\circle*{0.8}}
\put(  15.57,   23.15){\circle*{0.8}}
\put(  15.87,   22.75){\circle*{0.8}}
\put(  16.18,   22.35){\circle*{0.8}}
\put(  16.51,   21.95){\circle*{0.8}}
\put(  16.84,   21.55){\circle*{0.8}}
\put(  17.19,   21.15){\circle*{0.8}}
\put(  17.55,   20.75){\circle*{0.8}}
\put(  17.93,   20.35){\circle*{0.8}}
\put(  18.32,   19.94){\circle*{0.8}}
\put(  18.72,   19.55){\circle*{0.8}}
\put(  19.12,   19.17){\circle*{0.8}}
\put(  19.52,   18.80){\circle*{0.8}}
\put(  19.92,   18.45){\circle*{0.8}}
\put(  20.32,   18.11){\circle*{0.8}}
\put(  20.72,   17.78){\circle*{0.8}}
\put(  21.12,   17.47){\circle*{0.8}}
\put(  21.52,   17.16){\circle*{0.8}}
\put(  21.92,   16.87){\circle*{0.8}}
\put(  22.32,   16.59){\circle*{0.8}}
\put(  22.72,   16.31){\circle*{0.8}}
\put(  23.12,   16.05){\circle*{0.8}}
\put(  23.52,   15.79){\circle*{0.8}}
\put(  23.92,   15.54){\circle*{0.8}}
\put(  24.32,   15.30){\circle*{0.8}}
\put(  24.72,   15.07){\circle*{0.8}}
\put(  25.12,   14.84){\circle*{0.8}}
\put(  25.52,   14.62){\circle*{0.8}}
\put(  25.92,   14.40){\circle*{0.8}}
\put(  26.32,   14.20){\circle*{0.8}}
\put(  26.72,   13.99){\circle*{0.8}}
\put(  27.12,   13.80){\circle*{0.8}}
\put(  27.52,   13.61){\circle*{0.8}}
\put(  27.92,   13.42){\circle*{0.8}}
\put(  28.32,   13.24){\circle*{0.8}}
\put(  28.72,   13.06){\circle*{0.8}}
\put(  29.12,   12.89){\circle*{0.8}}
\put(  29.52,   12.73){\circle*{0.8}}
\put(  29.92,   12.56){\circle*{0.8}}
\put(  30.32,   12.41){\circle*{0.8}}
\put(  30.72,   12.25){\circle*{0.8}}
\put(  31.12,   12.10){\circle*{0.8}}
\put(  31.52,   11.95){\circle*{0.8}}
\put(  31.92,   11.81){\circle*{0.8}}
\put(  32.32,   11.67){\circle*{0.8}}
\put(  32.72,   11.53){\circle*{0.8}}
\put(  33.13,   11.40){\circle*{0.8}}
\put(  33.53,   11.27){\circle*{0.8}}
\put(  33.93,   11.14){\circle*{0.8}}
\put(  34.33,   11.01){\circle*{0.8}}
\put(  34.73,   10.89){\circle*{0.8}}
\put(  35.13,   10.77){\circle*{0.8}}
\put(  35.53,   10.65){\circle*{0.8}}
\put(  35.93,   10.54){\circle*{0.8}}
\put(  36.33,   10.43){\circle*{0.8}}
\put(  36.73,   10.32){\circle*{0.8}}
\put(  37.14,   10.21){\circle*{0.8}}
\put(  37.54,   10.11){\circle*{0.8}}
\put(  37.94,   10.00){\circle*{0.8}}
\put(  38.34,    9.90){\circle*{0.8}}
\put(  38.74,    9.80){\circle*{0.8}}
\put(  39.14,    9.71){\circle*{0.8}}
\put(  39.54,    9.61){\circle*{0.8}}
\put(  39.94,    9.52){\circle*{0.8}}
\put(  40.34,    9.43){\circle*{0.8}}
\put(  40.74,    9.34){\circle*{0.8}}
\put(  41.14,    9.25){\circle*{0.8}}
\put(  41.55,    9.16){\circle*{0.8}}
\put(  41.95,    9.08){\circle*{0.8}}
\put(  42.35,    8.99){\circle*{0.8}}
\put(  42.75,    8.91){\circle*{0.8}}
\put(  43.15,    8.83){\circle*{0.8}}
\put(  43.55,    8.75){\circle*{0.8}}
\put(  43.95,    8.68){\circle*{0.8}}
\put(  44.35,    8.60){\circle*{0.8}}
\put(  44.75,    8.52){\circle*{0.8}}
\put(  45.15,    8.45){\circle*{0.8}}
\put(  45.56,    8.38){\circle*{0.8}}
\put(  45.96,    8.31){\circle*{0.8}}
\put(  46.36,    8.24){\circle*{0.8}}
\put(  46.76,    8.17){\circle*{0.8}}
\put(  47.16,    8.10){\circle*{0.8}}
\put(  47.56,    8.03){\circle*{0.8}}
\put(  47.96,    7.97){\circle*{0.8}}
\put(  48.36,    7.91){\circle*{0.8}}
\put(  48.76,    7.84){\circle*{0.8}}
\put(  49.16,    7.78){\circle*{0.8}}
\put(  49.57,    7.72){\circle*{0.8}}
\put(  49.97,    7.66){\circle*{0.8}}
\put(  50.37,    7.60){\circle*{0.8}}
\put(  50.77,    7.54){\circle*{0.8}}
\put(  51.17,    7.48){\circle*{0.8}}
\put(  51.57,    7.43){\circle*{0.8}}
\put(  51.97,    7.37){\circle*{0.8}}
\put(  52.37,    7.32){\circle*{0.8}}
\put(  52.77,    7.26){\circle*{0.8}}
\put(  53.17,    7.21){\circle*{0.8}}
\put(  53.58,    7.15){\circle*{0.8}}
\put(  53.98,    7.10){\circle*{0.8}}
\put(  54.38,    7.05){\circle*{0.8}}
\put(  54.78,    7.00){\circle*{0.8}}
\put(  55.18,    6.95){\circle*{0.8}}
\put(  55.58,    6.90){\circle*{0.8}}
\put(  55.98,    6.85){\circle*{0.8}}
\put(  56.38,    6.81){\circle*{0.8}}
\put(  56.78,    6.76){\circle*{0.8}}
\put(  57.18,    6.71){\circle*{0.8}}
\put(  57.59,    6.67){\circle*{0.8}}
\put(  57.99,    6.62){\circle*{0.8}}
\put(  58.39,    6.58){\circle*{0.8}}
\put(  58.79,    6.53){\circle*{0.8}}
\put(  59.19,    6.49){\circle*{0.8}}
\put(  59.59,    6.45){\circle*{0.8}}
\put(  59.99,    6.41){\circle*{0.8}}
\put(  60.39,    6.36){\circle*{0.8}}
\put(  60.79,    6.32){\circle*{0.8}}
\put(  61.19,    6.28){\circle*{0.8}}
\put(  61.60,    6.24){\circle*{0.8}}
\put(  62.00,    6.20){\circle*{0.8}}
\put(  62.40,    6.16){\circle*{0.8}}
\put(  62.80,    6.13){\circle*{0.8}}
\put(  63.20,    6.09){\circle*{0.8}}
\put(  63.60,    6.05){\circle*{0.8}}
\put(  64.00,    6.01){\circle*{0.8}}
\put(  64.40,    5.98){\circle*{0.8}}
\put(  64.80,    5.94){\circle*{0.8}}
\put(  65.20,    5.90){\circle*{0.8}}
\put(  65.60,    5.87){\circle*{0.8}}
\put(  66.00,    5.83){\circle*{0.8}}
\put(  66.40,    5.80){\circle*{0.8}}
\put(  66.80,    5.77){\circle*{0.8}}
\put(  67.20,    5.73){\circle*{0.8}}
\put(  67.60,    5.70){\circle*{0.8}}
\put(  68.00,    5.67){\circle*{0.8}}
\put(  68.40,    5.63){\circle*{0.8}}
\put(  68.80,    5.60){\circle*{0.8}}
\put(  69.20,    5.57){\circle*{0.8}}
\put(  69.60,    5.54){\circle*{0.8}}
\put(  70.00,    5.51){\circle*{0.8}}
\put(  70.40,    5.48){\circle*{0.8}}
\put(  70.80,    5.45){\circle*{0.8}}
\put(  71.20,    5.42){\circle*{0.8}}
\put(  71.60,    5.39){\circle*{0.8}}
\put(  72.00,    5.36){\circle*{0.8}}
\put(  72.40,    5.33){\circle*{0.8}}
\put(  72.80,    5.30){\circle*{0.8}}
\put(  73.20,    5.27){\circle*{0.8}}
\put(  73.60,    5.24){\circle*{0.8}}
\put(  74.00,    5.21){\circle*{0.8}}
\put(  74.40,    5.19){\circle*{0.8}}
\put(  74.80,    5.16){\circle*{0.8}}
\put(  75.20,    5.13){\circle*{0.8}}
\put(  75.60,    5.11){\circle*{0.8}}
\put(  76.00,    5.08){\circle*{0.8}}
\put(  76.40,    5.05){\circle*{0.8}}
\put(  76.80,    5.03){\circle*{0.8}}
\put(  77.20,    5.00){\circle*{0.8}}
\put(  77.60,    4.98){\circle*{0.8}}
\put(  78.00,    4.95){\circle*{0.8}}
\put(  78.40,    4.93){\circle*{0.8}}
\put(  78.80,    4.90){\circle*{0.8}}
\put(  79.20,    4.88){\circle*{0.8}}
\put(  79.60,    4.85){\circle*{0.8}}
\put(  80.00,    4.83){\circle*{0.8}}
\put(  80.40,    4.81){\circle*{0.8}}
\put(  80.80,    4.78){\circle*{0.8}}
\put(  81.20,    4.76){\circle*{0.8}}
\put(  81.60,    4.74){\circle*{0.8}}
\put(  82.00,    4.71){\circle*{0.8}}
\put(  82.40,    4.69){\circle*{0.8}}
\put(  82.80,    4.67){\circle*{0.8}}
\put(  83.20,    4.65){\circle*{0.8}}
\put(  83.60,    4.63){\circle*{0.8}}
\put(  84.00,    4.60){\circle*{0.8}}
\put(  84.40,    4.58){\circle*{0.8}}
\put(  84.80,    4.56){\circle*{0.8}}
\put(  85.20,    4.54){\circle*{0.8}}
\put(  85.60,    4.52){\circle*{0.8}}
\put(  86.00,    4.50){\circle*{0.8}}
\put(  86.40,    4.48){\circle*{0.8}}
\put(  86.80,    4.46){\circle*{0.8}}
\put(  87.20,    4.44){\circle*{0.8}}
\put(  87.60,    4.42){\circle*{0.8}}
\put(  88.00,    4.40){\circle*{0.8}}
\put(  88.40,    4.38){\circle*{0.8}}
\put(  88.80,    4.36){\circle*{0.8}}
\put(  89.20,    4.34){\circle*{0.8}}
\put(  89.60,    4.32){\circle*{0.8}}
\put(  90.00,    4.30){\circle*{0.8}}
\put(  90.40,    4.28){\circle*{0.8}}
\put(  90.80,    4.26){\circle*{0.8}}
\put(  91.20,    4.25){\circle*{0.8}}
\put(  91.60,    4.23){\circle*{0.8}}
\put(  92.00,    4.21){\circle*{0.8}}
\put(  92.40,    4.19){\circle*{0.8}}
\put(  92.80,    4.17){\circle*{0.8}}
\put(  93.20,    4.16){\circle*{0.8}}
\put(  93.60,    4.14){\circle*{0.8}}
\put(  94.00,    4.12){\circle*{0.8}}
\put(  94.40,    4.10){\circle*{0.8}}
\put(  94.80,    4.09){\circle*{0.8}}
\put(  95.20,    4.07){\circle*{0.8}}
\put(  95.60,    4.05){\circle*{0.8}}
\put(  96.00,    4.04){\circle*{0.8}}
\put(  96.40,    4.02){\circle*{0.8}}
\put(  96.80,    4.00){\circle*{0.8}}
\put(  97.20,    3.99){\circle*{0.8}}
\put(  97.60,    3.97){\circle*{0.8}}
\put(  98.00,    3.95){\circle*{0.8}}
\put(  98.40,    3.94){\circle*{0.8}}
\put(  98.80,    3.92){\circle*{0.8}}
\put(  99.20,    3.91){\circle*{0.8}}
\put(  99.60,    3.89){\circle*{0.8}}

\put(   0.03,   11.01){\circle*{0.8}}
\put(   0.43,   11.34){\circle*{0.8}}
\put(   0.71,   11.74){\circle*{0.8}}
\put(   0.91,   12.14){\circle*{0.8}}
\put(   1.07,   12.54){\circle*{0.8}}
\put(   1.21,   12.94){\circle*{0.8}}
\put(   1.34,   13.34){\circle*{0.8}}
\put(   1.45,   13.74){\circle*{0.8}}
\put(   1.56,   14.14){\circle*{0.8}}
\put(   1.67,   14.54){\circle*{0.8}}
\put(   1.76,   14.94){\circle*{0.8}}
\put(   1.86,   15.34){\circle*{0.8}}
\put(   1.95,   15.74){\circle*{0.8}}
\put(   2.03,   16.14){\circle*{0.8}}
\put(   2.11,   16.54){\circle*{0.8}}
\put(   2.19,   16.94){\circle*{0.8}}
\put(   2.27,   17.34){\circle*{0.8}}
\put(   2.35,   17.74){\circle*{0.8}}
\put(   2.42,   18.14){\circle*{0.8}}
\put(   2.49,   18.54){\circle*{0.8}}
\put(   2.56,   18.94){\circle*{0.8}}
\put(   2.63,   19.34){\circle*{0.8}}
\put(   2.70,   19.74){\circle*{0.8}}
\put(   2.77,   20.14){\circle*{0.8}}
\put(   2.83,   20.54){\circle*{0.8}}
\put(   2.90,   20.94){\circle*{0.8}}
\put(   2.96,   21.34){\circle*{0.8}}
\put(   3.03,   21.74){\circle*{0.8}}
\put(   3.09,   22.14){\circle*{0.8}}
\put(   3.15,   22.54){\circle*{0.8}}
\put(   3.21,   22.94){\circle*{0.8}}
\put(   3.27,   23.34){\circle*{0.8}}
\put(   3.33,   23.74){\circle*{0.8}}
\put(   3.39,   24.14){\circle*{0.8}}
\put(   3.44,   24.54){\circle*{0.8}}
\put(   3.50,   24.94){\circle*{0.8}}
\put(   3.56,   25.34){\circle*{0.8}}
\put(   3.62,   25.74){\circle*{0.8}}
\put(   3.67,   26.14){\circle*{0.8}}
\put(   3.73,   26.54){\circle*{0.8}}
\put(   3.78,   26.94){\circle*{0.8}}
\put(   3.84,   27.34){\circle*{0.8}}
\put(   3.89,   27.74){\circle*{0.8}}
\put(   3.95,   28.14){\circle*{0.8}}
\put(   4.00,   28.54){\circle*{0.8}}
\put(   4.05,   28.94){\circle*{0.8}}
\put(   4.11,   29.34){\circle*{0.8}}
\put(   4.16,   29.74){\circle*{0.8}}
\put(   4.21,   30.14){\circle*{0.8}}
\put(   4.27,   30.54){\circle*{0.8}}
\put(   4.32,   30.94){\circle*{0.8}}
\put(   4.37,   31.34){\circle*{0.8}}
\put(   4.42,   31.74){\circle*{0.8}}
\put(   4.48,   32.14){\circle*{0.8}}
\put(   4.53,   32.54){\circle*{0.8}}
\put(   4.58,   32.94){\circle*{0.8}}
\put(   4.63,   33.34){\circle*{0.8}}
\put(   4.68,   33.74){\circle*{0.8}}
\put(   4.73,   34.14){\circle*{0.8}}
\put(   4.79,   34.54){\circle*{0.8}}
\put(   4.84,   34.94){\circle*{0.8}}
\put(   4.89,   35.34){\circle*{0.8}}
\put(   4.94,   35.74){\circle*{0.8}}
\put(   4.99,   36.14){\circle*{0.8}}
\put(   5.04,   36.54){\circle*{0.8}}
\put(   5.09,   36.94){\circle*{0.8}}
\put(   5.14,   37.34){\circle*{0.8}}
\put(   5.20,   37.74){\circle*{0.8}}
\put(   5.25,   38.14){\circle*{0.8}}
\put(   5.30,   38.54){\circle*{0.8}}
\put(   5.35,   38.94){\circle*{0.8}}
\put(   5.40,   39.34){\circle*{0.8}}
\put(   5.45,   39.74){\circle*{0.8}}
\put(   5.50,   40.14){\circle*{0.8}}
\put(   5.56,   40.54){\circle*{0.8}}
\put(   5.61,   40.94){\circle*{0.8}}
\put(   5.66,   41.34){\circle*{0.8}}
\put(   5.71,   41.74){\circle*{0.8}}
\put(   5.76,   42.14){\circle*{0.8}}
\put(   5.81,   42.54){\circle*{0.8}}
\put(   5.87,   42.94){\circle*{0.8}}
\put(   5.92,   43.34){\circle*{0.8}}
\put(   5.97,   43.74){\circle*{0.8}}
\put(   6.02,   44.14){\circle*{0.8}}
\put(   6.08,   44.54){\circle*{0.8}}
\put(   6.13,   44.94){\circle*{0.8}}
\put(   6.18,   45.34){\circle*{0.8}}
\put(   6.23,   45.74){\circle*{0.8}}
\put(   6.29,   46.14){\circle*{0.8}}
\put(   6.34,   46.54){\circle*{0.8}}
\put(   6.39,   46.94){\circle*{0.8}}
\put(   6.45,   47.34){\circle*{0.8}}
\put(   6.50,   47.74){\circle*{0.8}}
\put(   6.56,   48.14){\circle*{0.8}}
\put(   6.61,   48.54){\circle*{0.8}}
\put(   6.67,   48.94){\circle*{0.8}}
\put(   6.72,   49.34){\circle*{0.8}}
\put(   6.78,   49.74){\circle*{0.8}}
\put(   6.83,   50.14){\circle*{0.8}}
\put(   6.89,   50.54){\circle*{0.8}}
\put(   6.95,   50.94){\circle*{0.8}}
\put(   7.00,   51.34){\circle*{0.8}}
\put(   7.06,   51.74){\circle*{0.8}}
\put(   7.12,   52.14){\circle*{0.8}}
\put(   7.17,   52.54){\circle*{0.8}}
\put(   7.23,   52.94){\circle*{0.8}}
\put(   7.29,   53.34){\circle*{0.8}}
\put(   7.35,   53.74){\circle*{0.8}}
\put(   7.41,   54.14){\circle*{0.8}}
\put(   7.47,   54.54){\circle*{0.8}}
\put(   7.53,   54.94){\circle*{0.8}}
\put(   7.59,   55.34){\circle*{0.8}}
\put(   7.65,   55.74){\circle*{0.8}}
\put(   7.71,   56.14){\circle*{0.8}}
\put(   7.77,   56.54){\circle*{0.8}}
\put(   7.84,   56.94){\circle*{0.8}}
\put(   7.90,   57.34){\circle*{0.8}}
\put(   7.96,   57.74){\circle*{0.8}}
\put(   8.03,   58.14){\circle*{0.8}}
\put(   8.09,   58.54){\circle*{0.8}}
\put(   8.16,   58.94){\circle*{0.8}}
\put(   8.23,   59.34){\circle*{0.8}}
\put(   8.29,   59.74){\circle*{0.8}}
\put(   8.36,   60.14){\circle*{0.8}}
\put(   8.43,   60.54){\circle*{0.8}}
\put(   8.50,   60.94){\circle*{0.8}}
\put(   8.57,   61.34){\circle*{0.8}}
\put(   8.64,   61.74){\circle*{0.8}}
\put(   8.71,   62.14){\circle*{0.8}}
\put(   8.78,   62.54){\circle*{0.8}}
\put(   8.86,   62.94){\circle*{0.8}}
\put(   8.93,   63.34){\circle*{0.8}}
\put(   9.01,   63.74){\circle*{0.8}}
\put(   9.08,   64.14){\circle*{0.8}}
\put(   9.16,   64.54){\circle*{0.8}}
\put(   9.24,   64.94){\circle*{0.8}}
\put(   9.32,   65.34){\circle*{0.8}}
\put(   9.40,   65.74){\circle*{0.8}}
\put(   9.48,   66.14){\circle*{0.8}}
\put(   9.56,   66.54){\circle*{0.8}}
\put(   9.64,   66.94){\circle*{0.8}}
\put(   9.73,   67.34){\circle*{0.8}}
\put(   9.82,   67.74){\circle*{0.8}}
\put(   9.90,   68.14){\circle*{0.8}}
\put(   9.99,   68.54){\circle*{0.8}}
\put(  10.09,   68.94){\circle*{0.8}}
\put(  10.18,   69.34){\circle*{0.8}}
\put(  10.27,   69.74){\circle*{0.8}}
\put(  10.37,   70.14){\circle*{0.8}}
\put(  10.46,   70.54){\circle*{0.8}}
\put(  10.56,   70.94){\circle*{0.8}}
\put(  10.67,   71.34){\circle*{0.8}}
\put(  10.77,   71.74){\circle*{0.8}}
\put(  10.87,   72.14){\circle*{0.8}}
\put(  10.98,   72.54){\circle*{0.8}}
\put(  11.09,   72.94){\circle*{0.8}}
\put(  11.20,   73.35){\circle*{0.8}}
\put(  11.32,   73.75){\circle*{0.8}}
\put(  11.43,   74.15){\circle*{0.8}}
\put(  11.55,   74.55){\circle*{0.8}}
\put(  11.67,   74.95){\circle*{0.8}}
\put(  11.80,   75.35){\circle*{0.8}}
\put(  11.93,   75.75){\circle*{0.8}}
\put(  12.06,   76.15){\circle*{0.8}}
\put(  12.19,   76.55){\circle*{0.8}}
\put(  12.33,   76.95){\circle*{0.8}}
\put(  12.47,   77.35){\circle*{0.8}}
\put(  12.61,   77.75){\circle*{0.8}}
\put(  12.76,   78.15){\circle*{0.8}}
\put(  12.91,   78.55){\circle*{0.8}}
\put(  13.07,   78.95){\circle*{0.8}}
\put(  13.23,   79.35){\circle*{0.8}}
\put(  13.40,   79.75){\circle*{0.8}}
\put(  13.57,   80.15){\circle*{0.8}}
\put(  13.74,   80.55){\circle*{0.8}}
\put(  13.92,   80.95){\circle*{0.8}}
\put(  14.11,   81.35){\circle*{0.8}}
\put(  14.30,   81.75){\circle*{0.8}}
\put(  14.50,   82.15){\circle*{0.8}}
\put(  14.70,   82.55){\circle*{0.8}}
\put(  14.91,   82.95){\circle*{0.8}}
\put(  15.13,   83.35){\circle*{0.8}}
\put(  15.36,   83.75){\circle*{0.8}}
\put(  15.59,   84.15){\circle*{0.8}}
\put(  15.83,   84.55){\circle*{0.8}}
\put(  16.08,   84.95){\circle*{0.8}}
\put(  16.35,   85.35){\circle*{0.8}}
\put(  16.62,   85.75){\circle*{0.8}}
\put(  16.90,   86.15){\circle*{0.8}}
\put(  17.20,   86.55){\circle*{0.8}}
\put(  17.51,   86.95){\circle*{0.8}}
\put(  17.83,   87.35){\circle*{0.8}}
\put(  18.17,   87.75){\circle*{0.8}}
\put(  18.53,   88.15){\circle*{0.8}}
\put(  18.90,   88.55){\circle*{0.8}}
\put(  19.30,   88.95){\circle*{0.8}}
\put(  19.70,   89.33){\circle*{0.8}}
\put(  20.10,   89.70){\circle*{0.8}}
\put(  20.50,   90.04){\circle*{0.8}}
\put(  20.90,   90.37){\circle*{0.8}}
\put(  21.30,   90.69){\circle*{0.8}}
\put(  21.70,   90.99){\circle*{0.8}}
\put(  22.10,   91.27){\circle*{0.8}}
\put(  22.50,   91.54){\circle*{0.8}}
\put(  22.90,   91.80){\circle*{0.8}}
\put(  23.30,   92.05){\circle*{0.8}}
\put(  23.70,   92.29){\circle*{0.8}}
\put(  24.10,   92.52){\circle*{0.8}}
\put(  24.50,   92.73){\circle*{0.8}}
\put(  24.90,   92.94){\circle*{0.8}}
\put(  25.30,   93.14){\circle*{0.8}}
\put(  25.70,   93.33){\circle*{0.8}}
\put(  26.10,   93.51){\circle*{0.8}}
\put(  26.50,   93.69){\circle*{0.8}}
\put(  26.90,   93.86){\circle*{0.8}}
\put(  27.30,   94.02){\circle*{0.8}}
\put(  27.70,   94.18){\circle*{0.8}}
\put(  28.10,   94.33){\circle*{0.8}}
\put(  28.50,   94.47){\circle*{0.8}}
\put(  28.90,   94.61){\circle*{0.8}}
\put(  29.30,   94.74){\circle*{0.8}}
\put(  29.70,   94.87){\circle*{0.8}}
\put(  30.10,   95.00){\circle*{0.8}}
\put(  30.50,   95.12){\circle*{0.8}}
\put(  30.90,   95.23){\circle*{0.8}}
\put(  31.30,   95.34){\circle*{0.8}}
\put(  31.70,   95.45){\circle*{0.8}}
\put(  32.10,   95.55){\circle*{0.8}}
\put(  32.50,   95.65){\circle*{0.8}}
\put(  32.90,   95.75){\circle*{0.8}}
\put(  33.30,   95.84){\circle*{0.8}}
\put(  33.70,   95.94){\circle*{0.8}}
\put(  34.10,   96.02){\circle*{0.8}}
\put(  34.50,   96.11){\circle*{0.8}}
\put(  34.90,   96.19){\circle*{0.8}}
\put(  35.30,   96.27){\circle*{0.8}}
\put(  35.70,   96.35){\circle*{0.8}}
\put(  36.10,   96.42){\circle*{0.8}}
\put(  36.50,   96.50){\circle*{0.8}}
\put(  36.90,   96.57){\circle*{0.8}}
\put(  37.30,   96.63){\circle*{0.8}}
\put(  37.70,   96.70){\circle*{0.8}}
\put(  38.10,   96.76){\circle*{0.8}}
\put(  38.50,   96.83){\circle*{0.8}}
\put(  38.90,   96.89){\circle*{0.8}}
\put(  39.30,   96.95){\circle*{0.8}}
\put(  39.70,   97.00){\circle*{0.8}}
\put(  40.10,   97.06){\circle*{0.8}}
\put(  40.50,   97.11){\circle*{0.8}}
\put(  40.90,   97.17){\circle*{0.8}}
\put(  41.30,   97.22){\circle*{0.8}}
\put(  41.70,   97.27){\circle*{0.8}}
\put(  42.10,   97.32){\circle*{0.8}}
\put(  42.50,   97.36){\circle*{0.8}}
\put(  42.90,   97.41){\circle*{0.8}}
\put(  43.30,   97.46){\circle*{0.8}}
\put(  43.70,   97.50){\circle*{0.8}}
\put(  44.10,   97.54){\circle*{0.8}}
\put(  44.50,   97.58){\circle*{0.8}}
\put(  44.90,   97.62){\circle*{0.8}}
\put(  45.30,   97.66){\circle*{0.8}}
\put(  45.70,   97.70){\circle*{0.8}}
\put(  46.10,   97.74){\circle*{0.8}}
\put(  46.50,   97.78){\circle*{0.8}}
\put(  46.90,   97.81){\circle*{0.8}}
\put(  47.30,   97.85){\circle*{0.8}}
\put(  47.70,   97.88){\circle*{0.8}}
\put(  48.10,   97.91){\circle*{0.8}}
\put(  48.50,   97.95){\circle*{0.8}}
\put(  48.90,   97.98){\circle*{0.8}}
\put(  49.30,   98.01){\circle*{0.8}}
\put(  49.70,   98.04){\circle*{0.8}}
\put(  50.10,   98.07){\circle*{0.8}}
\put(  50.50,   98.10){\circle*{0.8}}
\put(  50.90,   98.13){\circle*{0.8}}
\put(  51.30,   98.15){\circle*{0.8}}
\put(  51.70,   98.18){\circle*{0.8}}
\put(  52.10,   98.21){\circle*{0.8}}
\put(  52.50,   98.23){\circle*{0.8}}
\put(  52.90,   98.26){\circle*{0.8}}
\put(  53.30,   98.28){\circle*{0.8}}
\put(  53.70,   98.31){\circle*{0.8}}
\put(  54.10,   98.33){\circle*{0.8}}
\put(  54.50,   98.36){\circle*{0.8}}
\put(  54.90,   98.38){\circle*{0.8}}
\put(  55.30,   98.40){\circle*{0.8}}
\put(  55.70,   98.42){\circle*{0.8}}
\put(  56.10,   98.44){\circle*{0.8}}
\put(  56.50,   98.47){\circle*{0.8}}
\put(  56.90,   98.49){\circle*{0.8}}
\put(  57.30,   98.51){\circle*{0.8}}
\put(  57.70,   98.53){\circle*{0.8}}
\put(  58.10,   98.55){\circle*{0.8}}
\put(  58.50,   98.56){\circle*{0.8}}
\put(  58.90,   98.58){\circle*{0.8}}
\put(  59.30,   98.60){\circle*{0.8}}
\put(  59.70,   98.62){\circle*{0.8}}
\put(  60.10,   98.64){\circle*{0.8}}
\put(  60.50,   98.65){\circle*{0.8}}
\put(  60.90,   98.67){\circle*{0.8}}
\put(  61.30,   98.69){\circle*{0.8}}
\put(  61.70,   98.70){\circle*{0.8}}
\put(  62.10,   98.72){\circle*{0.8}}
\put(  62.50,   98.74){\circle*{0.8}}
\put(  62.90,   98.75){\circle*{0.8}}
\put(  63.30,   98.77){\circle*{0.8}}
\put(  63.70,   98.78){\circle*{0.8}}
\put(  64.10,   98.80){\circle*{0.8}}
\put(  64.50,   98.81){\circle*{0.8}}
\put(  64.90,   98.82){\circle*{0.8}}
\put(  65.30,   98.84){\circle*{0.8}}
\put(  65.70,   98.85){\circle*{0.8}}
\put(  66.10,   98.86){\circle*{0.8}}
\put(  66.50,   98.88){\circle*{0.8}}
\put(  66.90,   98.89){\circle*{0.8}}
\put(  67.30,   98.90){\circle*{0.8}}
\put(  67.70,   98.92){\circle*{0.8}}
\put(  68.10,   98.93){\circle*{0.8}}
\put(  68.50,   98.94){\circle*{0.8}}
\put(  68.90,   98.95){\circle*{0.8}}
\put(  69.30,   98.96){\circle*{0.8}}
\put(  69.70,   98.97){\circle*{0.8}}
\put(  70.10,   98.99){\circle*{0.8}}
\put(  70.50,   99.00){\circle*{0.8}}
\put(  70.90,   99.01){\circle*{0.8}}
\put(  71.30,   99.02){\circle*{0.8}}
\put(  71.70,   99.03){\circle*{0.8}}
\put(  72.10,   99.04){\circle*{0.8}}
\put(  72.50,   99.05){\circle*{0.8}}
\put(  72.90,   99.06){\circle*{0.8}}
\put(  73.30,   99.07){\circle*{0.8}}
\put(  73.70,   99.08){\circle*{0.8}}
\put(  74.10,   99.09){\circle*{0.8}}
\put(  74.50,   99.10){\circle*{0.8}}
\put(  74.90,   99.11){\circle*{0.8}}
\put(  75.30,   99.12){\circle*{0.8}}
\put(  75.70,   99.13){\circle*{0.8}}
\put(  76.10,   99.14){\circle*{0.8}}
\put(  76.50,   99.14){\circle*{0.8}}
\put(  76.90,   99.15){\circle*{0.8}}
\put(  77.30,   99.16){\circle*{0.8}}
\put(  77.70,   99.17){\circle*{0.8}}
\put(  78.10,   99.18){\circle*{0.8}}
\put(  78.50,   99.19){\circle*{0.8}}
\put(  78.90,   99.19){\circle*{0.8}}
\put(  79.30,   99.20){\circle*{0.8}}
\put(  79.70,   99.21){\circle*{0.8}}
\put(  80.10,   99.22){\circle*{0.8}}
\put(  80.50,   99.22){\circle*{0.8}}
\put(  80.90,   99.23){\circle*{0.8}}
\put(  81.30,   99.24){\circle*{0.8}}
\put(  81.70,   99.25){\circle*{0.8}}
\put(  82.10,   99.25){\circle*{0.8}}
\put(  82.50,   99.26){\circle*{0.8}}
\put(  82.90,   99.27){\circle*{0.8}}
\put(  83.30,   99.27){\circle*{0.8}}
\put(  83.70,   99.28){\circle*{0.8}}
\put(  84.10,   99.29){\circle*{0.8}}
\put(  84.50,   99.29){\circle*{0.8}}
\put(  84.90,   99.30){\circle*{0.8}}
\put(  85.30,   99.31){\circle*{0.8}}
\put(  85.70,   99.31){\circle*{0.8}}
\put(  86.10,   99.32){\circle*{0.8}}
\put(  86.50,   99.33){\circle*{0.8}}
\put(  86.90,   99.33){\circle*{0.8}}
\put(  87.30,   99.34){\circle*{0.8}}
\put(  87.70,   99.34){\circle*{0.8}}
\put(  88.10,   99.35){\circle*{0.8}}
\put(  88.50,   99.35){\circle*{0.8}}
\put(  88.90,   99.36){\circle*{0.8}}
\put(  89.30,   99.37){\circle*{0.8}}
\put(  89.70,   99.37){\circle*{0.8}}
\put(  90.10,   99.38){\circle*{0.8}}
\put(  90.50,   99.38){\circle*{0.8}}
\put(  90.90,   99.39){\circle*{0.8}}
\put(  91.30,   99.39){\circle*{0.8}}
\put(  91.70,   99.40){\circle*{0.8}}
\put(  92.10,   99.40){\circle*{0.8}}
\put(  92.50,   99.41){\circle*{0.8}}
\put(  92.90,   99.41){\circle*{0.8}}
\put(  93.30,   99.42){\circle*{0.8}}
\put(  93.70,   99.42){\circle*{0.8}}
\put(  94.10,   99.43){\circle*{0.8}}
\put(  94.50,   99.43){\circle*{0.8}}
\put(  94.90,   99.44){\circle*{0.8}}
\put(  95.30,   99.44){\circle*{0.8}}
\put(  95.70,   99.45){\circle*{0.8}}
\put(  96.10,   99.45){\circle*{0.8}}
\put(  96.50,   99.45){\circle*{0.8}}
\put(  96.90,   99.46){\circle*{0.8}}
\put(  97.30,   99.46){\circle*{0.8}}
\put(  97.70,   99.47){\circle*{0.8}}
\put(  98.10,   99.47){\circle*{0.8}}
\put(  98.50,   99.48){\circle*{0.8}}
\put(  98.90,   99.48){\circle*{0.8}}
\put(  99.30,   99.48){\circle*{0.8}}
\put(  99.70,   99.49){\circle*{0.8}}

\end{picture}
\end{center}
\vspace*{10 mm}

\caption{Graphs of the functions $A(x)$ and $A_{0}(x)$.}
\end{figure}
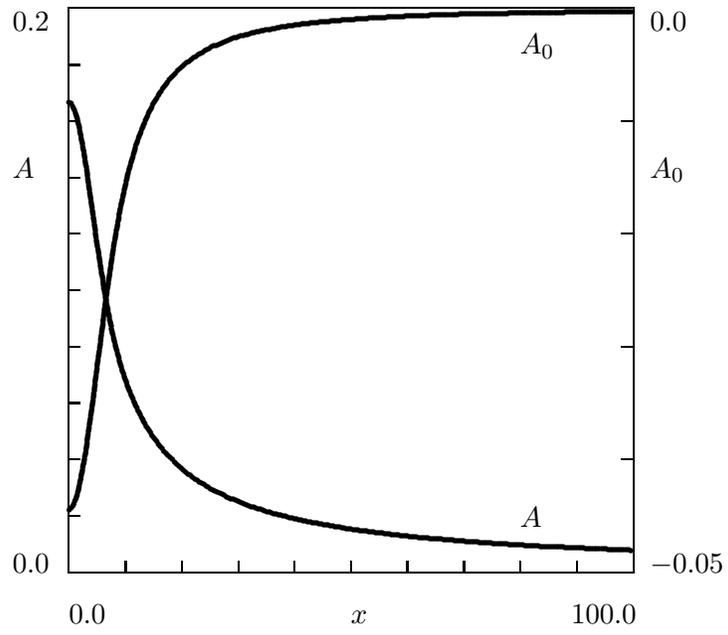

\end{document}